\documentclass[11pt,a4paper,english]{article}
\pdfoutput=1
\usepackage[T1]{fontenc}
\usepackage{subfigure}
\usepackage{graphicx}
\usepackage{amssymb}
\usepackage{esint}
\usepackage{a4wide}
\usepackage{psfrag}
\usepackage{babel}
\usepackage{epstopdf}
\makeatletter
\renewcommand{\theequation}{\hbox{\normalsize\arabic{section}.\arabic{equation}}}
\@addtoreset{equation}{section}
\renewcommand{\thefigure}{\hbox{\normalsize\arabic{section}.\arabic{figure}}}
\@addtoreset{figure}{section}
\renewcommand{\thetable}{\hbox{\normalsize\arabic{section}.\arabic{table}}}
\@addtoreset{table}{section}
\makeatother

\begin{document}
\begin{flushright}
{\normalsize SISSA 04/2009/EP \\
ITP-Budapest Report No. 643}
\end{flushright}

\vspace*{0.5cm}
\begin{center}
\LARGE
Effective potentials and kink spectra in \\
non-integrable perturbed conformal field theories
\end{center}
\vspace{0.5cm}
\begin{center}
{\large
G.\ Mussardo$^{1,2,3}$ and
G.\ Tak\'acs$^{4}$
\vspace{0.9cm}}

{\sl $^1$International School for Advanced Studies (SISSA),\\
Via Beirut 2-4, 34014 Trieste, Italy\\[2mm]
$^2$INFN Sezione di Trieste\\[2mm]
$^3$The Abdus Salam International Centre for Theoretical Physics, Trieste\\[2mm]
$^4$HAS Theoretical Physics Research Group, E\"otv\"os University \\
 P\'azm\'any P\'eter s\'et\'any 1/A, 1117 Budapest, Hungary}

\end{center}

\vspace{0.5cm}
\begin{center}
{\bf Abstract}\\[5mm]

\begin{tabular}{p{14cm}}
{\small
We analyze the evolution of the effective potential and the particle spectrum of two-parameter families of non-integrable quantum field theories. These theories are defined by deformations of conformal minimal models ${\mathcal M}_m$ by using the operators $\Phi_{1,3}$, $\Phi_{1,2}$ and $\Phi_{2,1}$. This study extends to all minimal models the analysis previously done for the classes of universality of the Ising, the Tricritical Ising and the RSOS models. We establish the symmetry and the duality properties of the various models also identifying the limiting theories that emerge when $m \rightarrow \infty$.
}
\end{tabular}

\end{center}

\newpage

\section{Introduction}
The aim of this paper is to understand in the easiest and most economical way the evolution of the effective potentials of particular non-integrable deformations of conformal field theories. We are interested, in particular, to analyze the relevant features of these effective potentials, such as 
the presence of stable and metastable vacua, the evolution of the spectrum of particles, the occurrence of confinement phenomena, etc. As shown below, quite a large amount of information can be gathered by using simple tools like the Form Factor Perturbation Theory, the Truncated Conformal Space Approach and the exact expressions of vacuum expectation values of relevant fields along the integrable directions. In the last two subjects -- quite crucial for our analysis -- one can easily perceive the unmistakable finger of Alyosha Zamolodchikov, a theoretical physicist who deeply changed the landscape of two-dimensional models, shaping it with his deep intuition, his profound knowledge of the field and his remarkable technical skills. This paper is warmly dedicated to him.  

\section{Basic aspects}\label{sec:intro}
The comprehension of the universality classes
of two-dimensional critical phenomena has been 
enormously enlarged in the last decades by the identification of the Conformal Field Theories of the fixed points \cite{BPZ} and the subsequent discovery of integrable deformations thereof \cite{Zam}. In this paper, we focus our attention, in particular, on the unitary conformal minimal models ${\mathcal M}_m$ ($m=3,4,\ldots$), with central charge and highest weights of the irreducible representations given by
\begin{equation}
c\,=\, 1-\frac{6}{m(m+1)}\,\hspace{5mm}
, \hspace{5mm} 
h_{r,s}\,=\, 
\frac{(r(m+1)-sm)^{2}-1}{4m(m+1)}\,=\, 
h_{m-r,m+1-s} \,\,\,.
\label{eq:kactable}
\end{equation}
A one-parameter family of deformed minimal models are defined by the action
\begin{equation}
\mathcal{A}_{m}^{(k,l)\pm}\,=\,
\mathcal{A}_{m}^{(CFT)} \pm \lambda\int dzd\bar{z}\Phi_{k,l}(z,\bar{z}) \,\,\,, 
\label{oneparameterfamily}
\end{equation}
where $\mathcal{A}_{m}$ is the action of the conformal minimal model, $\Phi_{r,s}$ is a relevant primary field with left/right conformal weights
$\Delta\,=\,\bar{\Delta}=h_{r,s}$ (with $\Delta < 1$)  and $\lambda>0$ is a dimensional 
coupling constant setting the scale of the quantum field theory\footnote{Note that the sign of the coupling only makes sense after fixing the
normalization of the fields $\Phi_{r,s}$, an issue to which we will return later. In terms of a mass scale $M$, $\lambda$ is expressed as $\lambda \sim M^{2(1 - h_{r,s})}$.}. Due to the null-vector structure of their Verma modules, the one-parameter deformations that generally define an integrable model away from criticality are those given by the relevant primary fields $\Phi_{1,3}$, $\Phi_{1,2}$ and $\Phi_{2,1}$ \cite{Zam}. The particle content of each of these integrable theories consists of kinks and/or bound states thereof, as we will shortly review in the next section. 

A two-parameter deformation of the conformal action ${\cal A}_m$ made of any pair of the fields $\Phi_{1,3}$, $\Phi_{1,2}$ and $\Phi_{2,1}$ leads however to a non-integrable model: in these cases it is impossible to find a matching of the null-vector structures of their Verma module able to define a set of conserved densities of higher spins \cite{GM}. In this paper we are interested in the analysis of the evolution of the effective potential by varying the coupling constants of the three non-integrable off-critical theories defined by 
\begin{eqnarray}
&& \mathcal{A}_{m}^{(1)}\,=\,
\mathcal{A}_{m}^{(CFT)} + \lambda_1 \int dzd\bar{z}\, \Phi_{1,3}(z,\bar{z}) + \mu_1 \int dzd\bar{z}\, \Phi_{1,2}(z,\bar{z}) \,\,\,; 
\nonumber \\
&& \mathcal{A}_{m}^{(2)}\,=\,
\mathcal{A}_{m}^{(CFT)} + \lambda_2 \int dzd\bar{z}\, \Phi_{1,3}(z,\bar{z}) + \mu_2 \int dzd\bar{z} \,\Phi_{2,1}(z,\bar{z}) \,\,\,;\\
&& \mathcal{A}_{m}^{(3)}\,=\,
\mathcal{A}_{m}^{(CFT)} + \lambda_3 \int dzd\bar{z} \,\Phi_{1,2}(z,\bar{z}) + \mu_3 \int dzd\bar{z} \,\Phi_{2,1}(z,\bar{z}) \,\,\,. \nonumber 
\label{twoparameterfamily}
\end{eqnarray}
In these three classes of non--integrable models, the two perturbations clearly play a symmetric role and each of them can be regarded as a deformation of the integrable theory defined by the other: for each theory, there is a dimensionless variable\footnote{In this formula $\Delta_{\lambda_i}$ and $\Delta_{\mu_i}$ denote the conformal weights of the conjugate fields to the coupling constants $\lambda_i$ and $\mu_i$ respectively. }  
\begin{equation}
\chi_i \equiv \lambda_i \,\mu_i^{-(1-
\Delta_{\lambda_i})/(1- \Delta_{\mu_i})} 
\,\,\,\,\,\,\,\,\,\,\,
,
\,\,\,\,\,\,\,\,\,\,\,
i=1,2,3
\label{rg}
\end{equation}
which characterizes two perturbative regimes:  the first is obtained in the limit $\chi_i \rightarrow 0$, while the other is reached for $\chi_i\rightarrow \infty$, that simply corresponds to swapping the role played by the two operators. The analytical control of the variation of the spectrum in both perturbative limits enables us to obtain interesting information about its evolution in the non-perturbative intermediate region.

Within the above classes of non-integrable models there are familiar examples of statistical models in their scaling limit. For instance, choosing $m=3$, the action ${\mathcal A}_3^{(1)}$ describes the full universality class of the Ising model that consists of the model displaced away from its critical temperature and in presence of an external magnetic field \cite{ffpt,FonsecaZam,DGM,Rutk}. Note that, for the symmetry of the conformal weights given in eqn.  (\ref{eq:kactable}), the universality class of the Ising model is also described by the action ${\mathcal A}_4^{(3)}$. For $m=4$, the action ${\mathcal A}_4^{(1)}$ describes instead the $Z_2$ spin even sector of the Tricritical Ising model, at a temperature different from its critical value and a non-critical value of the chemical potential of its vacancies \cite{LMT}. For a general value of $m$, the action ${\mathcal A}_m^{(1)}$ can be put in correspondence with deformations of the RSOS models and part of their phase space has been investigates in \cite{aldo}.  

A useful insight on the actions above, together with a bookkeeping of the $Z_2$ spin symmetry of the deformations, is provided by the Landau-Ginzburg (LG) formulation of the conformal unitary minimal models \cite{ZamLG}. In this approach, the conformal action ${\mathcal A}_m$ can be put in correspondence with the critical LG action of a scalar field $\varphi$, odd under the $Z_2$ symmetry 
$\varphi \rightarrow - \varphi$ 
\begin{equation}
{\mathcal A}_m \rightarrow {\mathcal A}_m^{(LG)}\,=\, \int dzd\bar{z}\,\left[
\frac{1}{2} (\partial \varphi)^2 + \varphi^{2 (m-1)} \right] \,\,\, , 
\label{criticalLG}
\end{equation}
The primary fields we are interested in are associated to the following normal ordered\footnote{Normal ordering is understood with respect to the conformal fusion rules of the primary fields of the model.} powers of the field $\varphi$
\begin{eqnarray}
\Phi_{1,3} & \rightarrow & : \varphi^{2 (m-2)} : \nonumber \\
\Phi_{1,2} & \rightarrow & : \varphi^{m-2}: \\
\Phi_{2,1} & \rightarrow & : \varphi^{m-1}: \nonumber 
\label{powersLG}
\end{eqnarray}
From these relations it follows that $\Phi_{1,2}$ and $\Phi_{2,1}$ have always a different $Z_2$ spin parity, independently of whether $m$ is an even or an odd number. The field $\Phi_{1,3}$ is always an even $Z_2$ field, while $\Phi_{1,2}$ ($\Phi_{2,1}$) is even (odd) only when $m$ is an even number and is an odd (even) $Z_2$ field otherwise. These observations will be useful in our analysis below.   

A generic feature of the non-integrable theories 
(\ref{twoparameterfamily}) consists of the confinement of the kinks of the original integrable models (associated to $\chi_i =0$ and $\chi_i=\infty$), accompanied by the appearance of new kink states and sequences of neutral bound states. The stability of these particles varies by varying $\chi_i$. It is important to estimate the spectrum of all these excitations, because the lowest kinks and the lowest neutral particles rule respectively the asymptotic behavior of the correlation functions in the topological and non-topological sectors of the theories. To investigate all these aspects, in the following we will mainly use the Form Factor Perturbation Theory \cite{ffpt,dsg}, further supported by certain continuity arguments. An explicit check of the validity of our theoretical conclusions can be provided by the numerical analysis coming from the Truncated Conformal Space Approach (TCSA) \cite{tcsa}, that can be implemented as follows. Consider the Hilbert space 
\begin{equation}
\mathcal{H}\,=\, \bigoplus_{(r,s)}\mathcal{V}_{r,s}\otimes\bar{\mathcal{V}}_{r,s}\label{eq:topzerospace}
\end{equation}
where $\mathcal{V}_{r,s}$ ($\bar{\mathcal{V}}_{r,s}$) denotes the irreducible representation of the left (right) Virasoro algebra with highest weight $h_{r,s}$, and the direct sum is modded out by the
$\mathbb{Z}_{2}$ symmetry of the Kac table (\ref{eq:kactable}) so that every different value of $h_{r,s}$ appears only once. In terms of conformal field theory, the decomposition (\ref{eq:topzerospace}) 
corresponds to the diagonal modular invariant partition function on a torus. Putting the theory in finite spatial volume $L$ by using the mapping
$
z\,=\,\mathrm{e}^{\frac{2\pi}{L}(\tau-ix)}
$ 
to pass from the conformal $z$-plane to the cylinder, with $x\sim x+L$,  we can define the off-critical Hamiltonian of the systems as 
\begin{equation}
H_{m}^{(i)} \,=\, H_m^{(CFT)} 
+ \sum_k V_k 
\label{eq:pcftham0}
\end{equation}
where 
\begin{equation}
H_m^{(CFT)} = \frac{2\pi}{L}\left(L_{0}+\bar{L}_{0}-\frac{c}{12}\right)
\,\,\,\,\,\,\,\, 
,
\,\,\,\,\,\,\,\,
V_k \,=\, g_k \int_{0}^{L} dx \,\Phi_k \,\,\,, 
\end{equation}
and we have used the shorthand notation $\Phi_k$ for the deforming fields and $g_k$ for the corresponding coupling.  $H_m^{(i)}$ is an infinite dimensional matrix in the space $\mathcal{H}$ (choosing an orthonormal basis of eigenvectors of $L_{0}$ and $\bar{L}_{0}$) and the matrix elements of its various terms are 
\begin{eqnarray}
&& \langle\Psi | H_{m}^{(0)} | \Psi' \rangle \, = \, \frac{2\pi}{L}\Bigg[\left(\Delta_{\Psi}+\bar{\Delta}_{\Psi}-\frac{c}{12}\right)\delta_{\Psi\Psi'}\Bigg]
\label{eq:pcftham}\\
&& \langle\Psi | V_k |\Psi'\rangle  \, =  \,
\frac{g_k L^{2-2h_{k}}}{(2\pi)^{1-2h_{k}}}\, 
\langle\Psi|\Phi_{k}(1,1)|\Psi'\rangle \, \delta_{s_{\Psi},s_{\Psi'}}
\nonumber 
\end{eqnarray}
where the matrix element of the perturbing operator is evaluated on the conformal plane at $z=\bar{z}=1$. Due to translational invariance,
the Hamiltonian is block-diagonal in the conformal spin
$
s_{\Psi}=\Delta_{\Psi}-\bar{\Delta}_{\Psi}
$
which is related to the spatial momentum: 
$
P=\frac{2\pi}{L}s_{\Psi}
$. 
The numerical diagonalization of $ H_m^{(i)}$ is then performed by truncating the conformal basis to a finite number of states. The matrix form of the off-critical Hamiltonians will be useful in our analysis of the deformed models. 

We plan to do the following: in Section 3 we review the one-parameter integrable deformations of the minimal models, focusing our attention on their particle content. In Section \ref{sec:nonint} we discuss the evolution of the spectrum of the two-parameter non-integrable theories, in Section 5 we analyze the limiting situations when $m\rightarrow \infty$, finally gathering  our conclusions in Section \ref{sec:conclusions}.

\section{One-parameter integrable theories}\label{sec:integrable}
This section provides an overview of known results on each individual integrable deformation. We start our analysis with the field $\Phi_{1,3}$, followed by the fields $\Phi_{1,2}$ and $\Phi_{2,1}$. A summary of the labels of the vacua and the main features of these theory can be found in Table \ref{tab.vacua} at the end of this section. 
 
\subsection{Kinks and vacua of the $\Phi_{1,3}$ perturbations}\label{subsec:13}

Consider the one-parameter deformation defined by the field $\Phi_{1,3}$. The fusion rules of this operator with the other primary fields of the theory 
\[
\left[\Phi_{1,3}\right]\times\left[\Phi_{r,s}\right]=\left[\Phi_{r,s-2}\right]+\left[\Phi_{r,s}\right]+\left[\Phi_{r,s+2}\right]
\]
imply that $\Phi_{1,3}$ only couples together subspaces $\mathcal{V}_{r,s}\otimes\bar{\mathcal{V}}_{r,s}$ and $\mathcal{V}_{r',s'}\otimes\bar{\mathcal{V}}_{r',s'}$ with $r=r'$ and $s-s'=0,\pm 2$. Therefore the Hilbert space (\ref{eq:topzerospace})
can be split further into $m-1$ separate sectors:
\begin{equation}
\mathcal{H}_{r}\,=\, 
\bigoplus_{{1\leq s\leq m\atop r+s\,\mathrm{even}}} \mathcal{V}_{r,s} \otimes 
\bar{\mathcal{V}}_{r,s}\quad,\quad r=1,\dots,m-1
\label{eq:13sectors}
\end{equation}
with the matrix elements of the perturbing operator that vanish between $\mathcal{H}_{r}$ and $\mathcal{H}_{r'}$ for $r\neq r'$. 
Choosing appropriate phase conventions for the fields, the theory $\mathcal{A}_{m}^{1,3(-)}$ has a massive spectrum, while the opposite sign of the perturbing operator leads to a massless flow whose endpoint is $\mathcal{A}_{m-1}$ in the infrared ($L\rightarrow\infty$) limit \cite{massless13}. The massive theory $\mathcal{A}_{m}^{1,3(-)}$
has $(m-1)$ degenerate vacua in infinite volume. Defining the theory on a finite volume of length $L$,  the degeneracy of the vacua is lifted by tunneling 
processes and the energy difference are split exponentially with $L$
\begin{equation}
E_{k}(L)-E_{l}(L)\sim\mathrm{e}^{-\mu L}
\label{eq:tunnelsplit}
\end{equation}
where the scale parameter $\mu$ is given by $
\mu\sim\lambda^{(m+1)/4}$, 
with a dimensionless number in front as the constant of proportionality. Therefore the finite volume splitting between the vacua is non-perturbative in the coupling $\lambda$.
\begin{figure}
\begin{centering}
\includegraphics[scale=1.3]{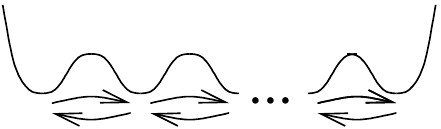}
\par\end{centering}

\caption{\label{fig:adj13}Adjacency rules and Landau-Ginzburg potential for
theories $\mathcal{A}_{m}^{1,3(-)}$}

\end{figure}

At the critical point $\lambda=0$, the ground state of the sector $\mathcal{H}_{r}$ is given by the state generated by the field $\Phi_{r,r}$, which also gives the ultraviolet ($L\rightarrow0$) limit of the $\lambda < 0$ (off-critical) ground state \cite{felderleclair}. Following the standard conventions, we label these vacua as $|{\bf a}\rangle$, where 
\[
{\bf a} =0,1/2,\dots,(m-2)/2 \,\,\,.
\]
The elementary excitations are kinks $| K_{ab} \rangle$ which interpolate between these ground states. They do not form bound state and are subjected to the adjacency rules
$
\left|\mathbf{a}-\mathbf{b}\right|=\frac{1}{2}.
$
This situation can be represented pictorially as shown in Figure \ref{fig:adj13}, where the continuous curve depicts a Landau-Ginzburg effective potential
with $m$ minima and the arrows indicate the interpolating kinks. These adjacency rules originate from the standard RSOS restriction based on a $U_{q}(sl(2))$ quantum group symmetry with 
\begin{equation}
q=\mathrm{e}^{i\pi\frac{m+1}{m}}\label{eq:q13}
\end{equation}
by associating the representation of spin ${\bf a}$ to the vacuum $|{\bf a}\rangle$ and putting the kinks in the spin $1/2$ representation. The tensor product rules of the quantum group give the following adjacency pattern
for the kinks: 
\begin{eqnarray*}
\frac{1}{2}\otimes0 & = & \frac{1}{2}\\
\frac{1}{2}\otimes {\bf a} & = & \left({\bf a} - \frac{1}{2} \right) \oplus \left({\bf a} + \frac{1}{2}\right) \quad {\bf a} > 0
\end{eqnarray*}
When $q$ is a root of unity as in (\ref{eq:q13}), RSOS restriction implements the truncation of the $U_{q}(sl(2))$ tensor product rules
as follows \cite{pasquier}:
\begin{equation}
{\bf a} \otimes {\bf b} \,=\, \bigoplus_{{\bf c} = |{\bf a} - {\bf b}|}^{\min({\bf a} + {\bf b},2j_{\mathrm{max}} - {\bf a} - 
{\bf b})} {\bf c} 
\label{eq:rsos_fusion}
\end{equation}
where 
$
j_{\mathrm{max}}=\frac{m-2}{2}
$
is the maximum spin allowed by the restriction. Note that the truncated tensor product rules (\ref{eq:rsos_fusion}) have the symmetry
\begin{equation}
{\bf a} \rightarrow j_{\mathrm{max}} - {\bf a} 
\label{eq:Z2}
\end{equation}
and, consequently, the adjacency graph is also symmetric. 

For later considerations, it is useful to introduce the quantum dimensions 
$\mathcal{D}_{\bf a}$ which satisfy
\begin{equation}
\mathcal{D}_{\bf a} \mathcal{D}_{\bf b} \,=\, \bigoplus_{{\bf c} = |{\bf a} - {\bf b}|}^{\min({\bf a} + {\bf b},2j_{\mathrm{max}} - {\bf a} - 
{\bf b})} \mathcal{D}_{\bf c} 
\label{eq:rsos_dimension_fusion}
\end{equation}
In the case of a group symmetry, when normalized to unity for the trivial
representation, the $\mathcal{D}$ are integers giving the dimensions of
the group representations. In the case of the truncated quantum group symmetry 
(\ref{eq:rsos_dimension_fusion}) can be solved to give
\begin{equation}
\mathcal{D}_{\bf a}=
\frac{ \sin\left(\frac{(2{\bf a}+1)\pi}{2(j_{\mathrm{max}}+1)}\right) }{\sin\left(\frac{\pi}{2(j_{\mathrm{max}}+1)}\right) }
\end{equation}
which are in general non-integer, reflecting the fact the the quantum symmetry
algebra underlying the kink structure is not a group.
In the case of $\mathcal{A}_{m}^{1,3(-)}$, the quantum dimension of the kink
multiplet turns out to be
\begin{equation}
\mathcal{D}_{1/2}=2\cos\left(\frac{\pi}{m}\right)
\label{kinkdeg13}\end{equation}
The physical meaning of this result can be seen as follows. The vacuum
energy of the model in a finite volume $R$ can be calculated using the 
NLIE description \cite{Dunning}:
\begin{equation}
E(R)=-2M\cos\left(\frac{\pi}{m}\right)\int_{-\infty}^{\infty}d\theta
\cosh\theta \mathrm{e}^{-MR\cosh\theta}
\end{equation}
where $M$ is the kink mass. This is exactly the contribution of a massive
multiplet of particles with the degeneracy given by (\ref{kinkdeg13}). For 
$m\rightarrow\infty$ $\mathcal{D}_{1/2}$ tends to $2$, 
which is in accordance with the
fact that the RSOS restriction is lifted and the model tends to the
sine-Gordon model in which the solitons form a doublet.

One can also look at the degeneracy of multi-kink states with given momenta,
i.e the number of possible states of the form
\begin{equation}
K_{{\bf a}_1{\bf a}_2}(\theta_1)K_{{\bf a}_2{\bf a}_3}(\theta_2)\dots K_{{\bf a}_n{\bf a}_{n+1}}(\theta_n)
\end{equation}  
where $K_{\bf ab}(\theta)$ denotes a kink of rapidity $\theta$ that connects
vacua ${\bf a}$ and ${\bf b}$. This is equivalent to enumerating the
allowed sequences
\begin{equation}
\mathcal{S}_n=\left\{({\bf a}_1,\dots,{\bf a}_n,{\bf a}_{n+1}):
\left|{\bf a}_i-{\bf a}_{i+1}\right|=\frac{1}{2};0\leq{\bf a}_i\leq j_{\mathrm
max} \right\}
\end{equation}
It turns out the number of allowed sequences grows as
$ 
%\begin{equation}
\mathrm{Card}\left(\mathcal{S}_n\right)\propto {D}_{1/2}^n
%\end{equation}
$. 
The $S$ matrix of the kinks, first obtained in \cite{reshetikhin_smirnov}, is also invariant under (\ref{eq:Z2}). For completeness we mention that the
$S$ matrix description of the massless flow $\mathcal{A}_{m}^{1,3(+)}$ is also known \cite{massless13Smat}.

\subsection{The kink spectrum for $\Phi_{1,2}$ perturbations}\label{subsec:12kinks}

\begin{figure}
\noindent \subfigure[$m$ odd]{\includegraphics[scale=1.3]{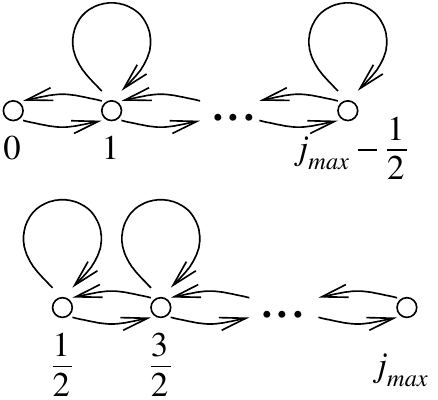}}\hfill\subfigure[$m$ even]{\includegraphics[scale=1.3]{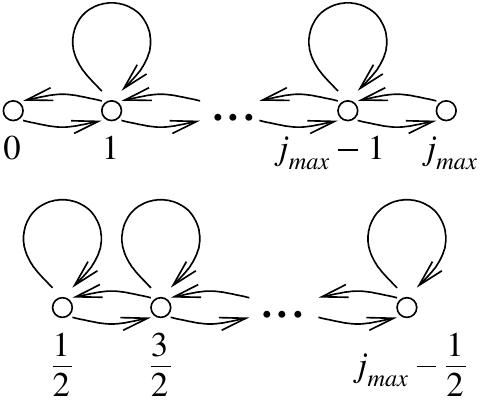}}

\caption{\label{fig:adj12} Adjacency rules for the kinks in $\Phi_{1,2}$
perturbations ($j_{\mathrm{max}}=\frac{m-2}{2}$)}

\end{figure}

For the models $\mathcal{A}_{m}^{(1,2)\pm}$ the conjectured $S$ matrix is based on a kink multiplet transforming in the spin $1$ representation of the same $U_{q}(sl(2))$ quantum group as specified
by (\ref{eq:q13}). Using the fusion rules (\ref{eq:rsos_fusion})
\begin{eqnarray*}
1\otimes0 & = & 1\\
1\otimes {\bf a} & = & ({\bf a} - 1) \oplus {\bf a} \oplus 
({\bf a} +1) 
\end{eqnarray*}
the adjacency graph now consists of two disconnected pieces, one containing
the vacua of integer spin and the other containing the vacua of half-integer spin. There are two separate possibilities depending on whether $m$
is odd or even, depicted in figure \ref{fig:adj12} (note that the fusion rule is again truncated by the RSOS restriction). 

The $S$-matrix of the kinks was obtained in \cite{Smat12} (see also \cite{rsos} for some corrections in the formulae for $6j$-symbols).
Apart from physically irrelevant phases (corresponding to redefinition of the phases of the vacua), it is again invariant under the symmetry
(\ref{eq:Z2}). However, there is a crucial difference: for even $m$ it leaves the two disconnected parts of the adjacency graph invariant, while for odd $m$ it swaps them. We will come back to this point later in this section. In the meantime let's recall that 
the full spectrum of the theory depends on the index $m$ \cite{Smat12}. In the three models of this deformation corresponding to $m=3,4,6$, there are only scalar particles: in these cases, in fact, also the kinks behave as scalar particles, and the spectra are associated to the root systems $E_8$, $E_7$ and $E_6$ respectively, with $8$, $7$ and $6$ particles \cite{Zam,MC,Sotkov,ZF}. For $m=5$ there are two degenerate kinks and one bound state thereof, degenerate with them. For $m > 6$, the spectrum consists instead of two kinks of mass 
\[
M \,\,\,,\,\,\,\,\,\,\,
2 M \cos\left(\frac{\pi}{3(m+1)}\right) 
\]
and two breathers of mass 
\[
2 M \sin\left(\frac{\pi m }{3(m+1)}\right)    
\,\,\,,\,\,\,\,\,\,\,
4 M \sin\left(\frac{\pi m}{3(m+1)}\right) 
\sin\left(\frac{\pi (2 m+1)}{3(m+1)}\right) 
\]
Their multiplicity depends on the vacuum structure that is examined in more detail below. Finally, for $m\rightarrow \infty$, the $\Phi_{1,2}$ perturbed model coincides with the usual sine-Gordon model, for a value of the frequency equal to $\beta^2 = 2 \pi$. At this value of the frequency, the spectrum of the sine-Gordon model consists of a breather of mass $M$ degenerate with the kink and the anti-kink, and an additional breather of mass $\sqrt{3} M$.

%\subsubsection{$\mathcal{A}_{m}^{(1,2)\pm}$ and $\mathcal{A}_{m}^{(2,1)\pm}$ as
%perturbed conformal field theories}\label{subsec:12pcft}

%\subsubsection{$\Phi_{1,2}$ perturbations}\label{subsubsec:pert12}
Let's now discuss the theories obtained by varying the sign of the coupling. First of all, in order to choose the sign of the coupling $\lambda$, we adopt the following conventions: all conformal two-point functions are assumed to be normalized as
\[
\left\langle \Phi_{r,s}(z,\bar{z})\Phi_{r',s'}(0,0)\right\rangle =\frac{\delta_{rr'}\delta_{ss'}}{z^{2h_{r,s}}\bar{z}^{2h_{r,s}}}
\]
while the conformal three-point couplings
\[
C_{(r',s')(r,s)}^{(r'',s'')}=\langle r'',s''|\Phi_{r',s'}(1,1)|r,s\rangle\quad\mbox{where}\quad|r,s\rangle=\Phi_{r,s}(0,0)|0\rangle
\]
are all real numbers due to unitarity. For $m$ even, the sign of the perturbing operator $\Phi_{1,2}$ can be fixed by demanding that
\begin{equation}
C_{(1,2)(k,k)}^{(k,k)} > 0\label{eq:sign_convention}
\end{equation}
This only makes sense because any field redefinition of $\Phi_{k,k}$ drops out. Note that there is no physical meaning to the sign of $\Phi_{1,2}$ when
$m$ is odd since the sign of every non-zero three-point couplings allowed by the fusion rules (\ref{eq:fusion12}) given below flips under the field redefinition 
\begin{equation}
\Phi_{r,s}\rightarrow(-1)^{s}\Phi_{r,s}
\label{eq:fieldredef}
\end{equation}
Let us start our discussion with the case when $m=2k+1$ is odd. The fusion rules of the perturbing operator $\Phi_{1,2}$
\begin{equation}
\left[\Phi_{1,2}\right]\times\left[\Phi_{r,s}\right]=\left[\Phi_{r,s-1}\right]+\left[\Phi_{r,s+1}\right]\label{eq:fusion12}
\end{equation}
imply that the independent sectors correspond to the following subspaces:
\begin{equation}
\mathcal{H}_{r}=\bigoplus_{s=1}^{m}\mathcal{V}_{r,s}\otimes\bar{\mathcal{V}}_{r,s}\quad,\quad r=1,\dots,k
\label{eq:12sectorsodd}
\end{equation}
There are then $k = (m-1)/2$ ground states, associated to the fields $\Phi_{r,r}$ of the  
sector $\mathcal{H}_{r}$. The adjacency rules in Figure \ref{fig:adj12} (a) fit this picture with either the set of integer or half-integer vacua (the choice is
irrelevant due to the symmetry (\ref{eq:Z2})). To fix the notation, for $m$ odd we choose as labels of the vacua the following values: 
\begin{eqnarray*}
\mathbf{a}\,=\,\frac{1}{2},\frac{3}{2},\ldots,\frac{m-2}{2} 
\,\,\,\,\,\,,\,\,\,\,\, \lambda > 0 \\
\mathbf{a}\,=\,0,1,\ldots,\frac{m-3}{2} 
\,\,\,\,\,\,,\,\,\,\,\, \lambda < 0
\end{eqnarray*} 
On the other hand, from the fusion rules (\ref{eq:fusion12}) the perturbing Hamiltonian in sector $\mathcal{H}_{r}$ has the following form: 
\[
\left(\begin{array}{ccccc}
0 & f_{1,2} & \dots & 0 & 0\\
f_{1,2}^{\dagger} & 0 & \dots & 0 & 0\\
\vdots & \vdots & \ddots & \vdots & \vdots\\
0 & 0 & \dots & 0 & f_{m-1,m}\\
0 & 0 & \dots & f_{m-1,m}^{\dagger} & 0\end{array}\right)
\]
where $f_{k,l}$ is the matrix of the operator $\Phi_{1,2}$ between the modules $\mathcal{V}_{r,k}$ and $\mathcal{V}_{r,l}$. Therefore it is obvious that the field redefinition (\ref{eq:fieldredef}) 
corresponds to a change $\lambda\rightarrow-\lambda$ in the Hamiltonian
(\ref{eq:pcftham}). As a result, for $m$ odd, the field theoretic models $\mathcal{A}_{m}^{(1,2)+}$
and $\mathcal{A}_{m}^{(1,2)-}$ are identical. This matches with the LG identification $\Phi_{1,2} \sim
\varphi^{m-2}$ that, for $m$ odd, predicts that $\Phi_{1,2}$ is an odd field under the $Z_2$ symmetry. 
It is also plausible that the state vector redefinition induced by (\ref{eq:fieldredef}) is just equivalent to the symmetry (\ref{eq:Z2}) which swaps the
integer and half-integer vacua. 
In comparison to the vacuum structure of $\Phi_{1,3}$ perturbations (where the sectors in eqn. (\ref{eq:13sectors}) were all independent), $\Phi_{1,2}$
induces a direct coupling (first order in $\lambda$) between the subspaces $\mathcal{H}_{r}$ and $\mathcal{H}_{m-r}$, since it couples
e.g. $\Phi_{r,r}$ and $\Phi_{m-r,m-r}=\Phi_{r,r+1}$. This explains why the number of ground states is half the number of those
obtained with a $\Phi_{13}$ perturbation, where all $\mathcal{H}_{r}$
are uncoupled. Here the coupling
between $\mathcal{H}_{r}$ and $\mathcal{H}_{m-r}$ is perturbative in $\lambda$ and grows with the volume, with the result to have a finite gap at $L\rightarrow\infty$. Note that the field redefinition
(\ref{eq:fieldredef}) just swaps the relative sign of the two subspaces $\mathcal{H}_{r}$ and $\mathcal{H}_{m-r}$.

For even $m$ let's pose $m=2k$, with the sectors now given by
\begin{eqnarray}
\mathcal{H}_{r} & = & \bigoplus_{s=1}^{m}\mathcal{V}_{r,s}\otimes\bar{\mathcal{V}}_{r,s}\quad,\quad r=1,\dots,k-1\nonumber \\
\mathcal{H}_{k} & = & \bigoplus_{s=1}^{k}\mathcal{V}_{k,s}\otimes\bar{\mathcal{V}}_{k,s}
\label{eq:12sectorseven}
\end{eqnarray}
The field redefinition (\ref{eq:fieldredef}) can still be used to establish that the sectors $\mathcal{H}_{r}$, $r=1,\dots,k-1$ are invariant under $\lambda\rightarrow-\lambda$. However, due to the Kac table symmetry 
\[
\left[\Phi_{k,k}\right] \,=\, \left[\Phi_{k,k+1}\right]
\]
we have 
\[
\left[\Phi_{1,2}\right]\times\left[\Phi_{k,k}\right]\,=\,
\left[\Phi_{k,k-1}\right]+\left[\Phi_{k,k}\right]
\]
and therefore the Hamiltonian matrix in sector $\mathcal{H}_{k}$ has a block along the diagonal:
\[
\left(\begin{array}{ccccc}
0 & f_{1,2} & \dots & 0 & 0\\
f_{1,2}^{\dagger} & 0 & \dots & 0 & 0\\
\vdots & \vdots & \ddots & \vdots & \vdots\\
0 & 0 & \dots & 0 & f_{k-1,k}\\
0 & 0 & \dots & f_{k-1,k}^{\dagger} & f_{k,k}\end{array}\right)
\]
Due to the presence of $f_{k,k}$, the spectrum in sector $\mathcal{H}_{k}$ is not invariant under $\lambda\rightarrow-\lambda$ and the models $\mathcal{A}_{m}^{(1,2)+}$ and $\mathcal{A}_{m}^{(1,2)-}$ are, in principle, different. This is, of course, in agreement with the LG position 
$\Phi_{1,2} \sim \varphi^{m-2}$ that, for $m$ even, states that $\Phi_{1,2}$ is even under the $Z_2$ symmetry. The number of vacua of $\mathcal{A}_{m}^{(1,2)+}$ is equal to $(m-2)/2$, while the number of vacua of $\mathcal{A}_{m}^{(1,2)-}$ is equal to $m/2$: to fix the notation, in this case we choose as labels of the vacua the following values: 
\begin{eqnarray*}
&& {\bf a}\,=\,\frac{1}{2},\frac{3}{2},\ldots,\frac{m-3}{2} 
\,\,\,\,\,\,,\,\,\,\,\, \lambda > 0 \\
&& {\bf a}\,=\,0,1,\ldots,\frac{m-2}{2} 
\,\,\,\,\,\,,\,\,\,\,\, \lambda < 0
\end{eqnarray*}
Such an assignment was previously fixed in \cite{exactvev} based on a conjecture for the exact vacuum expectation values of local primary
fields. A detailed evidence that models $\mathcal{A}_{2k}^{(1,2)+}$ are described by kink scattering theories built on vacua $|\bf{a}\rangle$ where $\bf{a}$ takes half-integer values, while $\mathcal{A}_{2k}^{(1,2)-}$ are associated to the kink scattering theories built on vacua $\{\bf{a}\rangle$, with $\bf{a}$ integer, is given in Appendix A using as example $m=6$.
 
For $m$ even, it is then natural to identify the theories $\mathcal{A}_m^{(1,2)\pm}$ with the two vacuum structures displayed in Figure \ref{fig:adj12} (b) and to assume that they
are related by duality.  This is in agreement with the known case of $m=4$ (the Tricritical Ising Model) 
\cite{LMT,LMC} and $m=6$ discussed in Appendix A. Note that, although the
naive counting of the kinks of the two models $\mathcal{A}_m^{(1,2)\pm}$ gives
different value, the dimensions of their Hilbert space are equal. This comes
from taking into proper account the composition law of the multi-kink
states. The quantum dimension of the kink multiplet in this case is given by 
\begin{equation}
\mathcal{D}_{1}=1+2\cos\left(\frac{2\pi}{m}\right)
\label{kinkdeg12}\end{equation}
which goes to $3$ when $m\rightarrow\infty$ corresponding to the fact that the
unrestricted soliton is a triplet \cite{Smat12}. Similarly to the case of the $\Phi_{1,2}$
perturbations the quantum dimension gives the growth of the degeneracy of multi-kink states.

%It is also interesting that the quantum
%dimensions of integer and half-integer vacua are related by the identity
%\begin{equation}
%\sum_{{\bf a}\ \mathrm{integer}}\mathcal{D}_{\bf a}^2=\sum_{{\bf
%    a}\ \mathrm{half-integer}}\mathcal{D}_{\bf
  %a}^2=\frac{m}{2\sin^2\frac{\pi }{m}}
%\end{equation}

%\vspace{1mm}

%\noindent \textbf{Proposition. }\emph{Models $\mathcal{A}_{2k}^{(1,2)+}$ are described by kink scattering theories built on vacua $|\bf{a}\rangle$
%where $\bf{a}$ taking half-integer values, while for $\mathcal{A}_{2k}^{(1,2)-}$ the relevant scattering amplitudes correspond to kinks supported by
%vacua with integer labels.}
In finite volume with periodic boundary conditions our previous considerations show that $\mathcal{A}_m^{(1,2)\pm}$ differ only in the sector $\mathcal{H}_k$ (where again $m=2k$). This means a very close relation between the multi-particle transfer matrices which determine the finite size spectrum. Using the Bethe-Yang description of multi-particle levels (valid when neglecting corrections exponentially decaying with the volume), the transfer matrix eigenvalues determining the states in the sectors $\mathcal{H}_r$ with $r=1,\dots,k-1$ must be exactly identical. Exact matching of the spectra in these sectors, however, places further constraints on the transfer matrix eigenvalues, since they enter the thermodynamic Bethe Ansatz equations which give an exact description of the finite size spectrum. For the spectra of the two models to agree in all sectors but one, all the eigenvalues (including also those corresponding to states in $\mathcal{H}_k$) must be nontrivially related. For $m=6$ the eigenvalues of the two-particle transfer matrix are given in eqns. (\ref{minusevals},\ref{plusevals}) and it is obvious that only two of them are functionally different:\[
\Lambda_{1}^{(+)}(\theta) = \Lambda_{1}^{(-)}(\theta)\quad,\quad \Lambda_{2}^{(+)}(\theta) =  \Lambda_{4}^{(-)}(\theta)\]
for either sign of the coupling. All the other eigenvalues can be expressed as the two functionally different ones multiplied by constant phase factors which correspond to twisted boundary conditions on the kinks. The allowed values of the twist depends on whether we consider the high-temperature ($\mathcal{A}_m^{(1,2)+}$) or low-temperature phase ($\mathcal{A}_m^{(1,2)-}$), which is a characteristic feature of low/high temperature (Kramers-Vannier) duality. This duality is well-known in the case $m=4$ (tricritical Ising) but it appears to be a more general feature of all the models $\mathcal{A}_m^{(1,2)\pm}$ with $m$ even. 
%This result is contrary to the discussion in \cite{exactvev} which states that duality is present only in the tricritical Ising model $m=4$.

\subsection{The kink spectrum for $\Phi_{2,1}$ perturbations}\label{subsec:21kinks}
%\subsubsection{$\Phi_{2,1}$ perturbations}

This case can be obtained from that of $\Phi_{1,2}$ perturbations by simply transposing the Kac table or, equivalently, exchanging the two relative prime numbers $p,p'$ labeling the minimal models (which
for unitary models satisfy $|p-p'|=1$) 
\cite{Smat12,exactvev}. For the model $\mathcal{A}_{m}^{(2,1)\pm}$ ($p=m$, $p'=m+1$) the quantum
group has the parameter
\begin{equation}
q=\mathrm{e}^{\pi i\frac{m}{m+1}}\label{eq:q21}
\end{equation}
In this case RSOS restriction dictates that $j_{\mathrm{max}} = (m-1)/2$ and so the adjacency graphs are the same as the ones for the model
$\mathcal{A}_{m+1}^{(1,2)\pm}$. This is in agreement with the LG position (\ref{powersLG}) 
of the fields $\Phi_{1,2}$ and $\Phi_{2,1}$.  
As a result, $\Phi_{2,1}$ perturbations are invariant under changing the sign of the couplings only when $m$ is even, while for $m$ odd the transformation $\lambda\rightarrow-\lambda$ corresponds to a change in the vacuum structure analogous to the case of $\mathcal{A}_{m}^{(1,2)\pm}$ when $m$ is even:  $\Phi_{2,1}$ is in fact odd under the $Z_2$ spin symmetry when $m$ is even (the theory is then  independent of the sign of its coupling), while is a $Z_2$ even field when $m$ is odd (and the theories 
with $\lambda >0$ and $\lambda < 0$ are related by duality). For the degenerate vacua, in this case we have the following labels   
\begin{itemize}
\item when $m$ is even, both for $\lambda > 0$ and $\lambda < 0$ their number is $m/2$ and 
\begin{eqnarray*}
&& \mathbf{a}\,=\,\frac{1}{2},\frac{3}{2},\ldots,\frac{m-1}{2} 
\,\,\,\,\,\,\,\,,\,\,\,\,\,\, \lambda > 0 \\
&& \mathbf{a}\,=\,0,1,\ldots,\frac{m-2}{2} 
\,\,\,\,\,\,\,\,\,\,,\,\,\,\,\,\, \,\,\lambda < 0
\end{eqnarray*}
\item when $m$ is odd, their number is $(m-1)/2$ for 
$\lambda >0$, and $(m+1)/2$ for $\lambda < 0$, with  
\begin{eqnarray*}
&& \mathbf{a}\,=\,\frac{1}{2},\frac{3}{2},\ldots,\frac{m-2}{2} 
\,\,\,\,\,\,\,,\,\,\,\,\, \lambda > 0 \\
&& \mathbf{a}\,=\,0,1,\ldots,\frac{m-1}{2} 
\,\,\,\,\,\,\,\,\,\,,\,\,\,\,\, \lambda < 0 
\end{eqnarray*}
\end{itemize} 
 
All the considerations about duality in the previous section carry over to the $\Phi_{2,1}$ case; the only difference is that the roles of even and odd $m$ models are interchanged, and nontrivial duality manifests itself for models with $m$ odd.

\begin{figure}
\begin{centering}
\subfigure[]{\includegraphics[bb=0bp 0bp 36bp 54bp,scale=1.3]{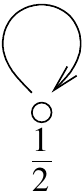}}~~~~~~~~~~\subfigure[]{\includegraphics[bb=0bp 0bp 36bp 30bp,scale=1.5]{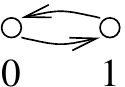}}
\par\end{centering}

\caption{\label{fig:ising}The two phases of the Ising model: (a) $T>T_{c}$
($\mathcal{A}_{3}^{(2,1)+}$) ~~(b) $T<T_{c}$ ($\mathcal{A}_{3}^{(2,1)-}$)}

\end{figure}

As an example, let us consider the simplest case $m=3$, which corresponds to the thermal perturbation of the Ising model. The relevant adjacency graphs are shown in Figure \ref{fig:ising} and they are in complete agreement with the fact that in the high-temperature phase the Ising
model has a single ground state supporting a bosonic particle with $S=-1$, while in the low-temperature phase there are two vacua interpolated
by a kink doublet corresponding to a Majorana fermion.

\begin{table}[t]
\begin{center}
\vspace{2mm}
\begin{tabular}{|c|c|c|c|}
\hline
Model & LG & Number of vacua & Vacuum index \\
\hline 
&&&\\ 
$\mathcal{A}_m^{(1,3)-}$ & $\varphi^{2(m-2)}$ & 
$m-1$ & $ a=0,\frac{1}{2},1,\ldots,\frac{m-2}{2}$ \\
&&&\\
\hline
&&&\\
& & $\frac{m-1}{2}$ ($m$ odd) & $ a= \frac{1}{2},\frac{3}{2},\ldots,\frac{m-2}{2}$ \\
$\mathcal{A}_m^{(1,2)+}$ & $\varphi^{m-2}$ & & \\
& &  $\frac{m-2}{2}$ ($m$ even) & $ a= \frac{1}{2},\frac{3}{2},\ldots,\frac{m-3}{2}$ \\
&&&\\
\hline
&&&\\
& & $\frac{m-1}{2}$ ($m$ odd) & $ a=0,1, \ldots,\frac{m-3}{2}$ \\
$\mathcal{A}_m^{(1,2)-}$ & $\varphi^{(m-2)}$ & & \\
& &  $\frac{m}{2}$ ($m$ even) & $ a= 0,1,\ldots,\frac{m-2}{2}$ \\
&&&\\ 
\hline
&&&\\
& & $\frac{m-1}{2}$, $m$ odd & $ a=\frac{1}{2},\frac{3}{2}, \ldots,\frac{m-2}{2}$ \\
$\mathcal{A}_m^{(2,1)+}$ & $\varphi^{m-1}$ & & \\
& &  $\frac{m}{2}$, $m$ even & $ a= \frac{1}{2},\frac{3}{2},\ldots,\frac{m-1}{2}$ \\
&&&\\
\hline 
&&&\\
& & $\frac{m+1}{2}$, $m$ odd & $ a=0,1,\ldots,\frac{m-1}{2}$ \\
$\mathcal{A}_m^{(2,1)-}$ & $\varphi^{m-1}$ & & \\
& &  $\frac{m}{2}$, $m$ even & $ a= 0,1,\ldots,\frac{m-2}{2}$ \\
&&&\\
\hline
\end{tabular}
\vspace{3mm}
\caption{\label{tab.vacua}Summary of the vacuum structure of the integrable models $\mathcal{A}_m^{(1,3)-}$, $\mathcal{A}_m^{(1,2)\pm}$ and 
$\mathcal{A}_m^{(2,1)\pm}$. 
}

\end{center}
\end{table}

\section{Non-integrable perturbations}\label{sec:nonint}

We now show how the picture presented in Section \ref{sec:integrable} changes when we consider non-integrable perturbations of minimal models obtained by adding two perturbing operators. Our arguments involve form factor perturbation theory developed in \cite{ffpt,dsg}.

\subsection{Perturbing with $\Phi_{1,3}$ and $\Phi_{1,2}$}\label{sec:1312}

Consider the theory given by the action
\[
\mathcal{A}_{m}^{(1)}\,=\,
\mathcal{A}_{m}^{(CFT)} + \lambda_1 \int dzd\bar{z}\, \Phi_{1,3}(z,\bar{z}) + \mu_1 \int dzd\bar{z}\, \Phi_{1,2}(z,\bar{z}) \,\,\,,
\]
characterized by the dimensionless coupling constant combination $\chi_1 = \lambda_1 \,\mu_1^{-(1-h_{1,3})/(1-h_{1,2})}$. This theory has been originally analyzed in \cite{aldo}.  As we are going to show below, for generic values of $\chi_1$, the kink spectrum of this theory is essentially the one inherited by the $\mathcal{A}_m^{(1,2)\pm}$ theory, with a discontinuity in the spectrum that occurs only for $\chi_1 \rightarrow \mp\infty$. Namely, in the plane of the couplings $\lambda_1$ and $\mu_1$ (see Figure \ref{vacuaevolution}), the numbers of degenerate vacua present in the half planes on the right and left hand sides of the vertical axis are generically equal to those of the $\mathcal{A}_m^{(1,2)\pm}$ theory: if $m$ is odd, there are 
$(m-1)/2$ vacua both in the right and left half planes, while if $m$ is even, there are  $(m-2)/2$ vacua in the half plane on the left side and $m/2$ 
in the half plane on the right side. In the former case, the two theories are identical, while in the latter case they are related by duality.  A different vacuum structure is only present in the limits $\chi_1 \rightarrow \mp \infty$: (i) when $\chi_1 \rightarrow - \infty$, there is a discontinuous jump in the number of vacua, that becomes equal to $(m-1)$, and for this reason $\chi_1=-\infty$ corresponds to a first order phase transition; (ii)  when $\chi_1 \rightarrow +\infty$, all vacua merge together, bringing the system to a massless phase, corresponding to the cross-over from the minimal models $\mathcal{M}_m \rightarrow 
\mathcal{M}_{m-1}$. 

The scenario discussed above can be inferred by 
studying the two perturbative regimes that can be approached by the form factor perturbation theory: $\chi_1 \rightarrow -\infty$ and $\chi_1 \rightarrow 0$. They are discussed separately below.

\subsubsection{$\mathcal{A}_{m}^{(1)}$ as the $\Phi_{1,2}$ perturbation of $\mathcal{A}_{m}^{(1,3)-}$}

\begin{figure}[t]\label{vacuashiftss}
\hspace{-20pt}
%\begin{centering} 
\subfigure{\includegraphics[scale=0.3]{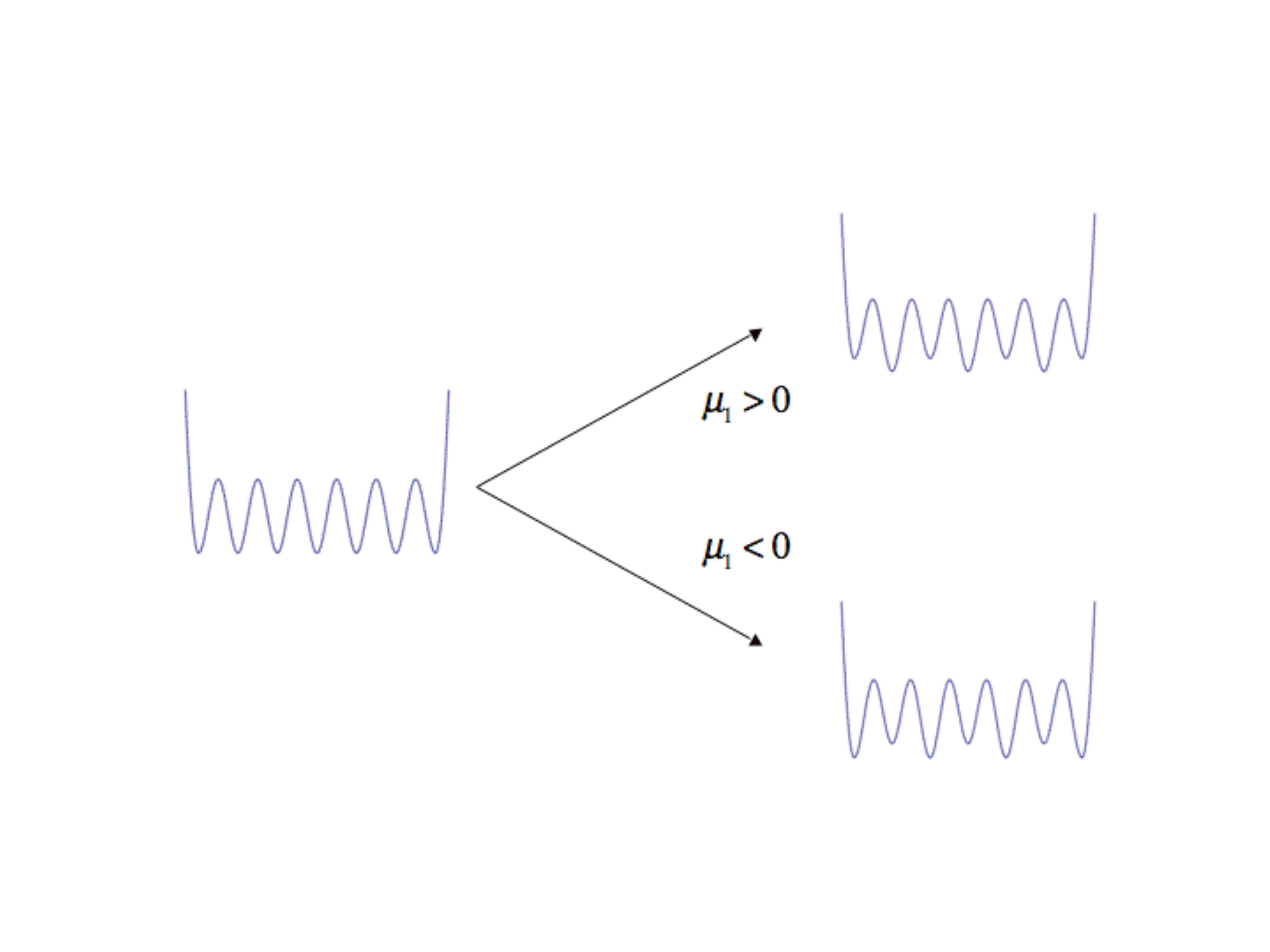}}
\subfigure{\includegraphics[scale=0.3]{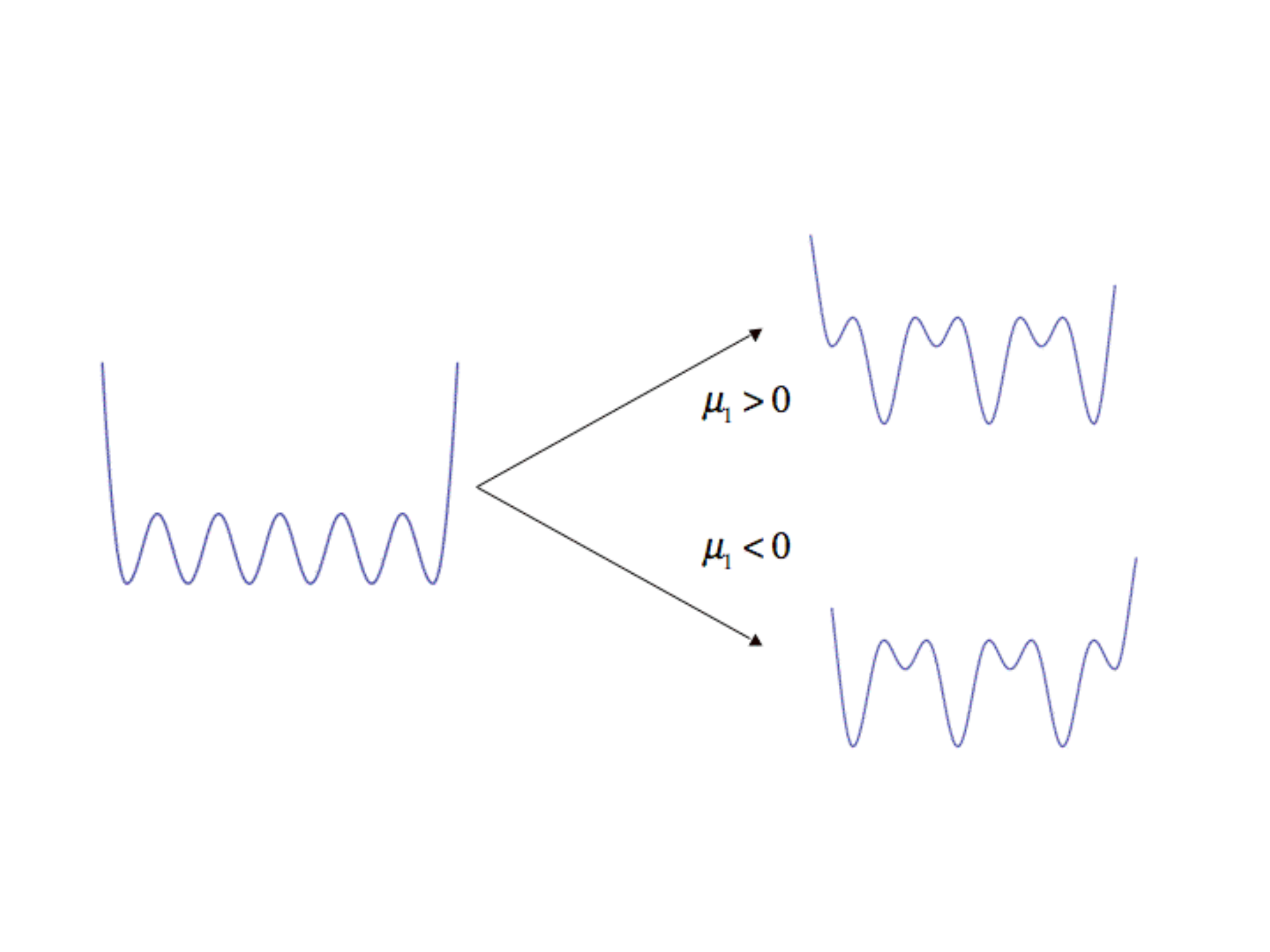}}
\par
%\end{centering}
\caption{{\em (a) $\Phi_{1,2}$ perturbation of the 
$\mathcal{A}_m^{(1,3)-}$ theory when $m$ is even (left-hand side) and when $m$ is odd (right-hand side). }}
\end{figure}

For $\lambda_1< 0$ and $\mu_1=0$ the theory is identical to the massive $\mathcal{A}_{m}^{(1,3)-}$ model, which has $(m-1)$ degenerate vacua
$| {\bf a}\rangle$ with $\mathbf{a}=0,1/2,\dots,m/2-1$. In the vicinity of $\chi_1 \rightarrow -\infty$, using the first order perturbation theory in $\mu_1$, the energy density of these vacua changes by the amount
\[
\delta\mathcal{E}_{\mathbf{a}} \,=\, \mu_1\langle {\bf a} |\Phi_{1,2}|{\bf a}\rangle^{(1,3)-}
\]
where the upper index indicates the model in which the matrix element is evaluated. The exact vacuum
expectation values of primary fields were conjectured in \cite{exactvev}:
\begin{equation}
\mathcal{A}_{m}^{(1,3)-}  :\quad  \langle {\bf a} |\Phi_{k,l}|{\bf a}\rangle^{(1,3)-}\, =\, 
\frac{\sin\left(\frac{\pi(2\mathbf{a}+1)}{m}((m+1)k-ml)\right)}{\sin\frac{\pi(2\mathbf{a}+1)}{m}}\, F_{k,l}^{m}(\lambda_1)\label{exactvev13}
\end{equation}
where the function $F^{m}_{k,l}$ is reported in Appendix B. In particular
\[
\langle {\bf a} | \Phi_{1,2} | {\bf a} \rangle^{(1,3)-}\, =\, 
(-1)^{2\mathbf{a}}\, F_{1,2}^{m}(\lambda_1) \,\,\,.
\]
So, the vacua with integer ${\bf a}$ receive
the same shift, while the ones with half-integer ${\bf a}$ are shifted by just the opposite amount. This supports the claim that, depending on the sign of the coupling, the perturbation $\Phi_{1,2}$ favors either the integer or the half-integer vacua (see Figure \ref{vacuashiftss}).
The perturbed theory nearby the negative vertical axis has therefore a new set of stable vacua (whose number coincides with the corresponding $\mathcal{A}_m^{(1,2)\pm}$ theory), while the remaining original ones become false vacua.  
%\begin{figure}
%\begin{centering}
%\psfrag{12}{$\frac{1}{2}$}\subfigure[]{\includegraphics[bb=0bp 0bp 36bp 54bp,scale=1.5]{isinghigh}}~~~~~~~~~~\psfrag{0}{$0$}\psfrag{1}{$1$}\subfigure[]{\includegraphics[bb=0bp 0bp 36bp 30bp,scale=1.5]{isinglow}}
%\par\end{centering}

%\caption{\label{fig:ising}The two phases of the Ising model: (a) $T>T_{c}$
%($\mathcal{A}_{3}^{(2,1)+}$) ~~(b) $T<T_{c}$ ($\mathcal{A}_{3}^{(2,1)-}$)}

%\end{figure}

%\begin{figure}[t]
%\begin{centering}
%\includegraphics[scale=0.5]{Slide2.pdf}
%\par\end{centering}
%\caption{\label{vacuashift} Structure of the vacua of the $\mu_1$ deformation of $\mathcal{A}_{m}^{(1,3)-}$, with $\mu_1 <0$ (upper figure) and $\mu_1 > 0$ (lower figure). This example refers to $m=8$}. 
%\end{figure}

Hence, in the perturbed theory the original kinks of $\mathcal{A}_{m}^{(1,3)-}$ disappear from the spectrum and they get confined. To determine the new spectrum of excitations, note that the perturbed theory has now a new set of degenerate vacua, corresponding either to integer of half-integer of ${\bf a}$, depending on the sign of $\mu_1$. 
Their number is either $(m-1)/2$ if $m$ is odd, or 
$m/2$ if $m$ is even: in both case it coincides with the number of vacua of the $A_m^{(1,2)\pm}$ theory. 
These vacua are connected by a new set of kinks 
$K'_{\mathbf{a},\mathbf{a}\pm1}$ that can be considered as bound states of original kinks of $\mathcal{A}_{m}^{(1,3)-}$
\cite{aldo}, of the form
\[
K'_{\mathbf{a},\mathbf{a}\pm 1} \sim K_{\mathbf{a},\mathbf{a}\pm\frac{1}{2}}K_{\mathbf{a}\pm\frac{1}{2},\mathbf{a}\pm1}
\]
If $M$ denotes the mass of the original kinks, at the lowest order in $\mu_1$ the mass of the new kink is $M' \simeq 2 M$. As we are going to show, this interpretation of the kink $K'$ as a bound states of the original kinks $K$ follows from form factor perturbation theory. Using the theory of superselection sectors presented in \cite{felderleclair}, the Hilbert space (\ref{eq:topzerospace}) can be identified in fact with
the sector of zero topological charge: periodic boundary conditions require that the multi-kink states 
\[
|K_{\mathbf{a}_{1}\mathbf{a}_{2}}(\theta_{1})K_{\mathbf{a}_{2}\mathbf{a}_{3}}(\theta_{2})\dots K_{\mathbf{a}_{n}\mathbf{a}_{n+1}}(\theta_{n})\rangle
\]
contained in that space satisfy $\mathbf{a}_{1} = \mathbf{a}_{n+1}$. The operators that create topologically charged sectors are nonlocal and they become chiral in the conformal limit. For the matrix elements of the Hamiltonian (\ref{eq:pcftham0}) to be well-defined,  they must be local with respect to the interaction potential, which is $\Phi_{1,3}$ for the model
$\mathcal{A}_{m}^{(1,3)-}$. The algebra of these operators is generated by chiral vertex operators $\varphi_{2,1}$ and $\bar{\varphi}_{2,1}$,
which can be considered as the ultraviolet limit of one-kink creation operators for the right/left movers respectively. Comparing with subsection
2.1, it is obvious that the fusion rules of the operators $\varphi_{k,1}$ (or equivalently $\bar{\varphi}_{k,1}$) are exactly identical to the tensor product decomposition (\ref{eq:rsos_fusion}) of the quantum
group representations truncated by the RSOS restriction, with $k=2\mathbf{a}+1$
giving the relation to the quantum group representation labeled by ${\bf a}$.
Under this identification $\varphi_{2,1}$ and $\bar{\varphi}_{2,1}$ transform in the doublet representation just as the elementary kinks
do.

The operators $\varphi_{2,1}$ and $\bar{\varphi}_{2,1}$ are, however, non-local with respect to the non-integrable perturbation $\Phi_{1,2}$
and, following an argument presented in \cite{dsg},  this results in an infinite mass correction at first order of form factor perturbation theory in $\mu$. This also means that the kink excitations corresponding to these operators are confined in the non-integrable theory. For two-kink states, however, the operator products
\begin{equation}
\varphi_{2,1}\varphi_{2,1}\sim\mathbb{I}+\varphi_{3,1}\quad,\quad\bar{\varphi}_{2,1}\bar{\varphi}_{2,1}\sim\mathbb{I}+\bar{\varphi}_{3,1}\quad,\quad\varphi_{2,1}\bar{\varphi}_{2,1}\sim\Phi_{2,1}\label{eq:chiralfusions}
\end{equation}
only contain operators that are local with respect to $\Phi_{1,2}$ and therefore two-kink bound states (either topologically charged or neutral) can
survive the perturbation.

In addition to the new set of kinks identified above, the perturbed theory has also neutral particles, while the unperturbed integrable model $\mathcal{A}_{m}^{(1,3)-}$ has none. In fact, the unbalance of the next-neighbor vacua of the original theory gives rise to a linear confinement potential between the kink and antikink of the original theory, with the consequent collapse of this pair into a string of bound states \cite{Mccoy-Wu,ffpt,dsg,FonsecaZam,Rutk}. The  tower of neutral bound states is the same for every new stable vacua of the perturbed theory. An estimate of their mass $m_k$ comes from semi-classical considerations: called $M$ the mass of the kink of the unperturbed $\mathcal{A}_{m}^{(1,3)-}$ theory and $F \equiv \mu_1 \,F_{1,2}^m(\lambda)$, one has \cite{Mccoy-Wu}
\begin{equation}
m_k \,=\, (2 + F^{2/3} \gamma_k^{2/3}) M \,\, ,
\label{chain}
\end{equation}
where $\gamma_k$ are the positive solutions of 
\[
J_{\frac{1}{3}} \left(\frac{1}{3} \gamma_k\right) + 
J_{-\frac{1}{3}} \left(\frac{1}{3} \gamma_k\right) = 0 \,\, ,
\]
with $J_{\nu}(x)$ being the Bessel function of order $\nu$. Not all of these particles are stable: the stable ones have to satisfy the condition $m_k < 2 m_1$ while those with $m_k > 2 m_1$, for the non-integrability of the theory, decay in the lower mass channels. The number of stable particles decreases by increasing the coupling $\mu_1$ and when $m_1$ reaches the threshold of $2 M'$ (where $M'$ is the mass of the kink of the perturbed theory) no neutral particles remain in the spectrum of the $\mathcal{A}_m^{(1)}$ theory.  

\subsubsection{$\mathcal{A}_{m}^{(1)}$ as the $\Phi_{1,3}$ perturbation
of $\mathcal{A}_{m}^{(1,2)\pm}$}

\begin{figure}[t]
\begin{centering}
\includegraphics[scale=0.4]{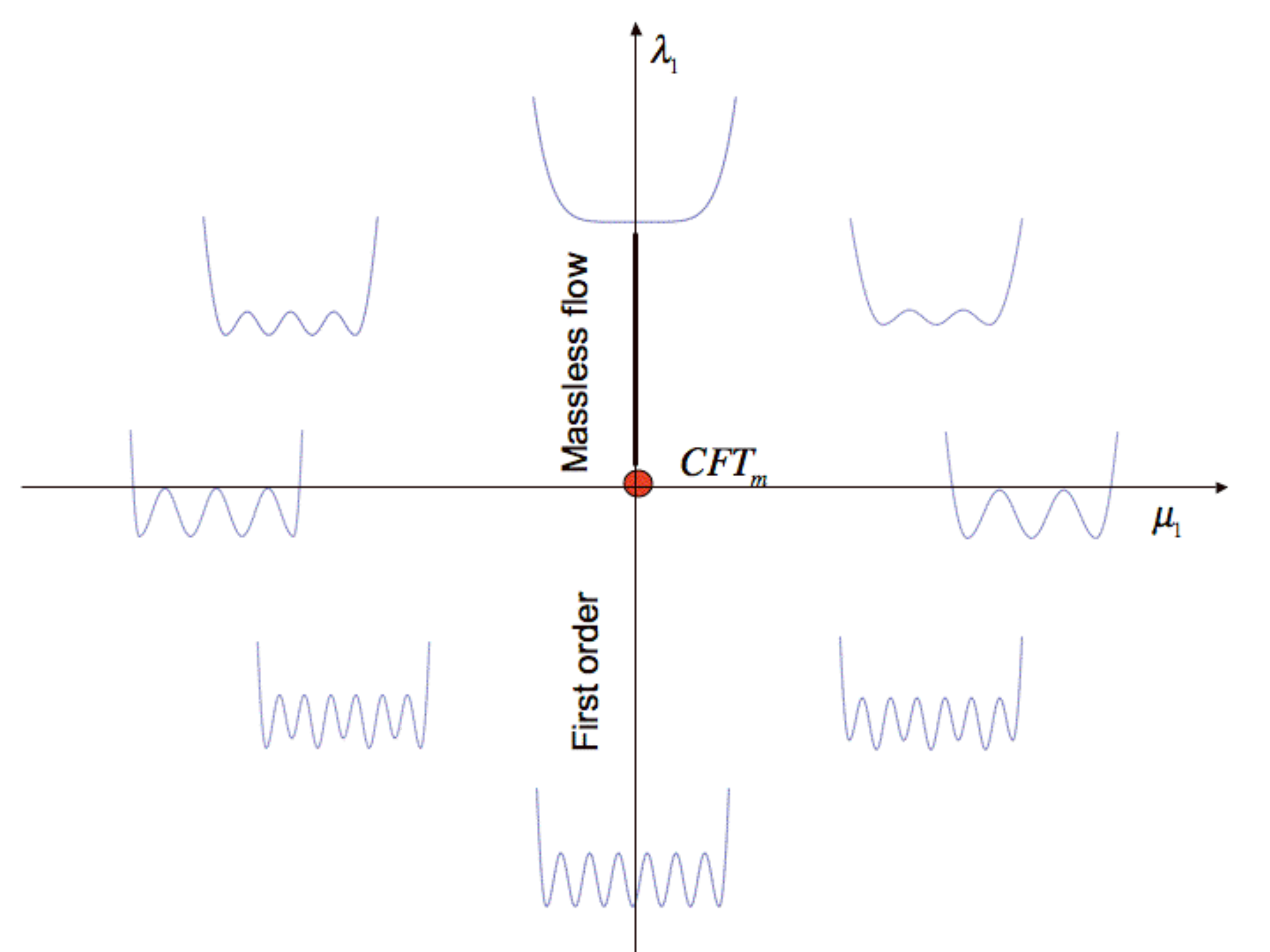}
\par\end{centering}

\caption{\label{vacuaevolution} Evolution of the vacua structure of the $\mathcal{A}_{m}^{(1)}$ theory in the plane of the couplings. This example refers to $m=8$.}
\end{figure}
Let's now consider the theory in the limit $\chi_1 \rightarrow 0$, where $\Phi_{1,3}$ can be regarded as the non-integrable perturbation added to an $\mathcal{A}_{m}^{(1,2)\pm}$ model. Unlike the other limit $\chi_1 \rightarrow -\infty$,  in this case the kink excitations of $\mathcal{A}_{m}^{(1,2)\pm}$ survive the perturbation and therefore they are present also for $\lambda \neq 0$. In fact, using the general formula of the vacuum expectation values \cite{exactvev}
\begin{equation}
\mathcal{A}_{m}^{(1,2)\pm}  :\quad  \langle {\bf a} | \Phi_{k,l} | {\bf a} \rangle^{(1,2)}\,=\, \frac{\sin\left(\frac{\pi(2\mathbf{a}+1)}{m}((m+1)k-ml)\right)}{\sin\frac{\pi(2\mathbf{a}+1)}{m}} \, 
G_{k,l}^{m}(\mu_1)
\label{exactvev12}
\end{equation}
(where $x=(m+1) k - m l$) with the function $G_{k,l}^m$ given in Appendix B, for the vacuum expectation value of $\Phi_{1,3}$ we have 
\[
\langle {\bf a} | \Phi_{1,3} | {\bf a} \rangle^{(1,2)}\, =\, - 
G_{1,3}^{m}(\mu_1) \,\,\,.
\]
Hence, all the vacua of $\mathcal{A}_{m}^{(1,2)\pm}$ are shifted by the {\em same amount} and their degeneracy is not broken by the perturbation. In addition, extending the considerations of the previous subsection for the $\Phi_{1,2}$ perturbations, the ultraviolet operators that creates the kinks are identified with $\varphi_{3,1}$ and $\bar{\varphi}_{3,1}$ (in accordance with subsection 2.2, these operators indeed correspond to the spin-$1$ representation of the quantum group). Both of them are local with respect to $\Phi_{1,3}$, so that the kinks of $\mathcal{A}_{m}^{(1,2)\pm}$ survive the $\Phi_{1,3}$ perturbation and, as shown by the first two fusions in (\ref{eq:chiralfusions}), they can indeed be considered as bound states of two $\mathcal{A}_{m}^{(1,3)-}$ kinks.

The above conclusions are also supported by considering the block structure of the perturbing Hamiltonian (\ref{eq:pcftham0}). The fusion rules operators of the operators $\Phi_{1,2}$ and $\Phi_{1,3}$ imply that the Hamiltonian still respects the block structure given in eqns. (\ref{eq:12sectorsodd}) and (\ref{eq:12sectorseven}). Therefore the vacuum structure is expected to coincide with that of the $\lambda=0$ theory, which is indeed what we found. These results are also consistent with the recent findings of \cite{LMT} in which the case of the tricritical Ising model ($m=4$) was considered. 
 
The evolution of the effective potential of the theory $\mathcal{A}_m^{(1)}$ is summarized in Figure \ref{vacuaevolution}, where the coupling constant $\mu_1$ and $\lambda_1$ are on the horizontal and on the vertical axis respectively. Note that moving from the horizontal axis toward the upper half plane, one expects that the heights of the maxima of the potential decrease so that, reaching the positive vertical there is a coalescence of all vacua, leading to a massless flow to the conformal model $\mathcal{M}_{m-1}$.    

%\vspace{3mm}
%{\bf WE SHALL DISCUSS THE THEORY THAT EMERGES WHEN $m \rightarrow \infty$, kind of 
%DOUBLE SINE-GORDON.}
%\vspace{3mm}

\subsection{Perturbing with $\Phi_{1,3}$ and $\Phi_{2,1}$}\label{subsec:1321}

The theory
\begin{equation}
\mathcal{A}_{m}^{(2)}\,=\,
\mathcal{A}_{m}^{(CFT)} + \lambda_2 \int dzd\bar{z}\, \Phi_{1,3}(z,\bar{z}) + \mu_2 \int dzd\bar{z} \,\Phi_{2,1}(z,\bar{z})
\label{eq:action1321}
\end{equation}
is characterized by the dimensionless combination of the coupling constants \[
\chi_2 = \lambda_2 \mu_2 ^{-(1-h_{1,3})/(1-h_{2,1})}\,\,\,.
\] 
As we are going to argue, for generic values of $\chi_2$, the model has either one vacuum (for $m$ even) or two degenerate vacua (for $m$ odd). In the plane of the couplings $\lambda_2$ and $\mu_2$, there are however certain lines of discontinuities: 
they correspond either to the values $\chi_2 \rightarrow \mp \infty$ or $\chi_2 \rightarrow 
0^{\pm}$. Concerning the first limits $\chi_2 \rightarrow \mp\infty$, the situation is as in the previous section: $\chi_2 \rightarrow -\infty$ corresponds to the first order phase transition of the $\mathcal{A}_m^{(1,3)-}$ theory, where there are $(m-1)$ vacua, while $\chi_2 \rightarrow +\infty$ corresponds to the massless flow $\mathcal{M}_m \rightarrow \mathcal{M}_{m-1}$. The vacuum structure of the two other limits 
$\chi_2 \rightarrow 0^{\mp}$ are instead described by the $\mathcal{M}_m^{(2,1)\pm}$ theories that, for $m$ even, both have $m/2$ vacua, whereas for $m$ odd, have $(m-1)/2$ ($\chi_2 \rightarrow 0^+$) and 
$(m+1)/2$ ($\chi_2 \rightarrow 0^-$) vacua respectively. These limits describe then first order phase transitions of the model in eqn. (\ref{eq:action1321}). 
 
Let's see how this scenario emerges by considering the evolution of the vacuum structure of the theory in the two perturbative limits $\chi_2 \rightarrow -\infty$ and $\chi_2 \rightarrow 0^{\pm}$.

\subsubsection{$\mathcal{A}_{m}^{(2)}$ as the $\Phi_{2,1}$ perturbation
of $\mathcal{A}_{m}^{(1,3)-}$}

Considering the model as a perturbation of $\mathcal{A}_{m}^{(1,3)-}$ (i.e. in the vicinity of 
$\chi_2 = -\infty$), the shifts of the energy of the vacua are ruled by the expectation values
\begin{equation}
\langle {\bf a} |\Phi_{2,1} |{\bf a} \rangle^{(1,3)}\,=\, 
(-1)^{2\mathbf{a}+1}2\cos\frac{\pi(2\mathbf{a}+1)}{m}\, F_{2,1}^{m}(\lambda_2)
\label{vac13pert21}\end{equation}
where 
\[
\mathbf{a}=0,\frac{1}{2},\dots,j_{\mathrm{max}}\quad,\quad j_{\mathrm{max}}=\frac{m-2}{2}\,\,\,.
\]
The expression of $F_{2,1}^m$ can be found in the Appendix B. There are two different possibilities depending on the parity of $m$.

\begin{figure}[t]
\begin{centering}
\includegraphics[scale=0.4]{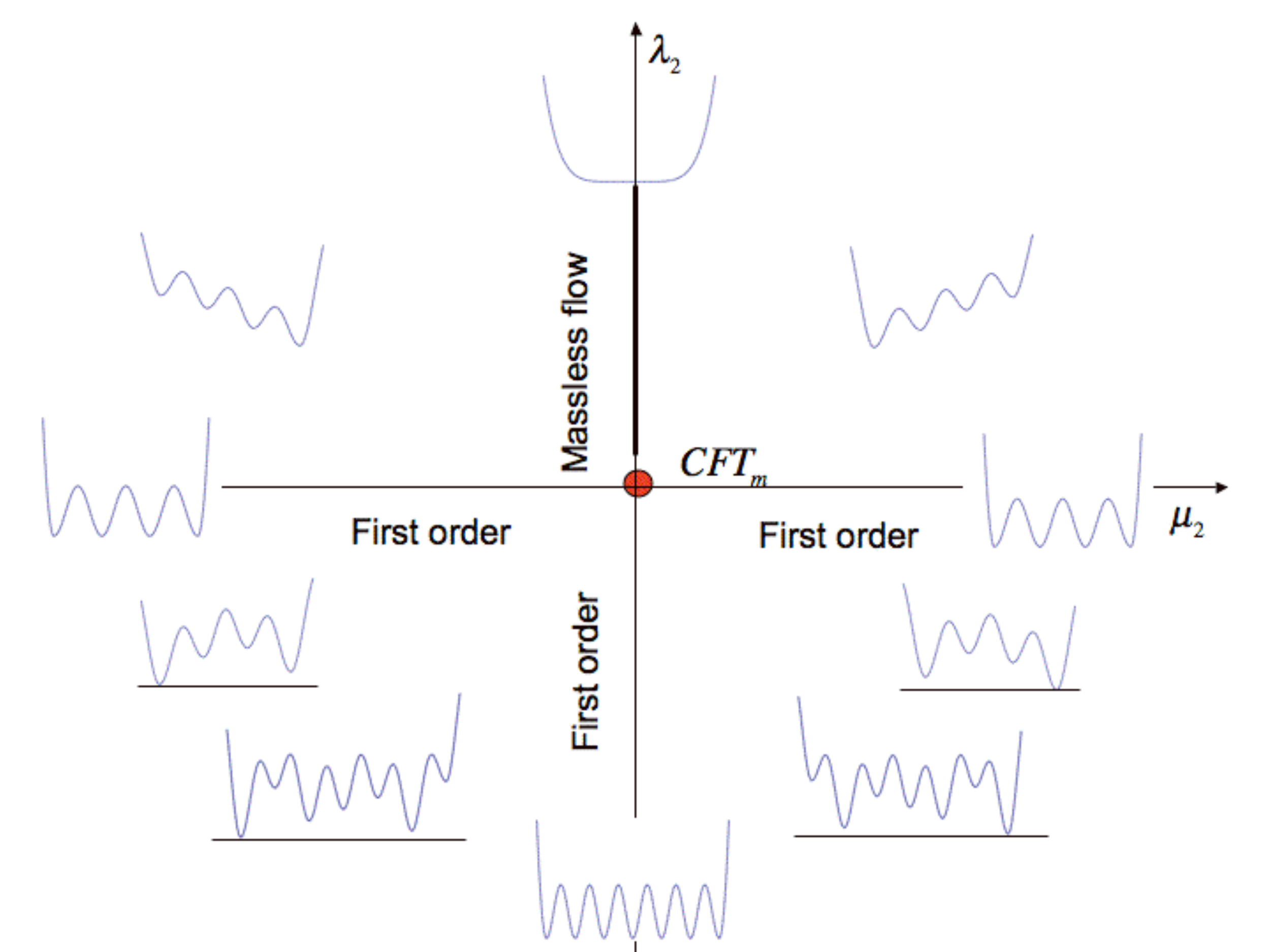}
\par\end{centering}

\caption{\label{vacuaevolutionn2} Evolution of the vacua structure of the $\mathcal{A}_{m}^{(2)}$ theory in the plane of the couplings when $m$ is even. This example refers to $m=8$.}
\end{figure}

\begin{itemize} 
\item 
When $m$ is even, all the vacuum degeneracies are not only lifted but also in a different manner, so that there is always a single vacuum left after switching on $\mu$. In agreement with this,  there are no topologically charged operators in this case 
that are local with respect to both perturbations, so all the original kinks are confined. The evolution of the vacua in this case is shown in Figure \ref{vacuaevolutionn2}.
\item When $m$ is odd, for the symmetry
\[
\langle j_{\mathrm{max}} - {\bf a} |\Phi_{2,1} | j_{\mathrm{max}} - {\bf a} \rangle^{(1,3)}\, =\, \langle {\bf a} |\Phi_{2,1} | {\bf a} \rangle^{(1,3)}
\]
the shifts of the vacua come in pairs but are otherwise nondegenerate. As a result, whatever sign is chosen for the coupling $\mu$, there are always two vacua left and all the others are shifted differently. Depending on the sign of the $\Phi_{2,1}$ coupling $\mu$, the two vacua are either the pair $\mathbf{a}=0$ and $\mathbf{a}=j_{\mathrm{max}}$ or the pair $\mathbf{a}=\frac{1}{2}$ and $\mathbf{a}=j_{\mathrm{max}}-\frac{1}{2}$. 

This can also be understood by considering the allowed topological charges. $\Phi_{1,3}$ is local with respect to the operators $\varphi_{k,1}$,
while $\Phi_{2,1}$ is local with respect to $\varphi_{1,2l+1}$. For odd $m$, the operator $\varphi_{m-1,1}\equiv\varphi_{1,m}$ is in the intersection of these two families, so the topological sector
created by it is still allowed. Therefore one expects a single kink doublet to be present in the spectrum, which interpolates between the two remaining vacua. It is easy to check that the allowed kink has the correct quantum numbers. Using the quantum group picture for $\Phi_{1,3}$, the operator $\varphi_{m-1,1}$ transforms in the spin
$j_{\mathrm{max}}$ representation. The RSOS truncated tensor product
rules (\ref{eq:rsos_fusion}) give the relations
\begin{eqnarray}
j_{\mathrm{max}}\otimes0=j_{\mathrm{max}}\quad & , & \quad j_{\mathrm{max}}\otimes j_{\mathrm{max}}=0\nonumber \\
j_{\mathrm{max}}\otimes\frac{1}{2}=j_{\mathrm{max}}-\frac{1}{2}\quad & , & \quad j_{\mathrm{max}}\otimes j_{\mathrm{max}}-\frac{1}{2}=\frac{1}{2}\label{eq:tprules21_13}
\end{eqnarray}
so the charged sectors created by $\varphi_{m-1,1}$ do indeed interpolate between the remaining two vacua. The simplest example is the Ising model $m=3$ where the two operators
are eventually identical $\Phi_{1,3}=\Phi_{2,1}$, and therefore the $\Phi_{2,1}$ perturbation just changes the scale of the $\mathcal{A}_{3}^{(1,3)-}$
but otherwise leaves the kink structure intact. 

\end{itemize}

\subsubsection{$\mathcal{A}_{m}^{(2)}$ as the $\Phi_{1,3}$ perturbation
of $\mathcal{A}_{m}^{(2,1)\pm}$}

We can also consider the model (\ref{eq:action1321}) as a $\Phi_{1,3}$ perturbation
of the $\mathcal{A}_{m}^{(2,1)\pm}$ model and studying it in the vicinity of $\chi_2 \rightarrow 0$. The exact vacuum expectation values conjectured in \cite{exactvev} have the form
\[
\langle {\bf a} | \Phi_{k,l} | {\bf a} \rangle^{(2,1)}\, =\, 
\frac{\sin\left(\frac{\pi(2\mathbf{a}+1)}{m+1}((m+1)k-ml)\right)}{\sin\frac{\pi(2\mathbf{a}+1)}{m+1}}\, H_{k,l}^{m}(\mu_2)
\]
where the function $H_{k,l}^m$ can be found in Appendix B. 
Specializing the above formula to $\Phi_{1,3}$ and simplifying it, we have
\begin{equation}
\langle {\bf a} | \Phi_{1,3} | {\bf a} \rangle^{(2,1)} \, =\, \frac{\sin\left(\frac{3\pi(2a+1)}{m+1}\right)}{\sin\frac{\pi(2a+1)}{m+1}}\, H_{1,3}^{m}(\mu_2)
\label{eq:13vevin21}
\end{equation}
Let us assume that the normalization of $\Phi_{1,3}$ is fixed so that $H_{1,3}^{m}(\mu_2) > 0$. Hence, the stable vacua are given by the maxima of the expression (\ref{eq:13vevin21}) for $\lambda < 0$,  and by the minima for $\lambda > 0$. In order to apply the above formula, one needs to refer to the proper values of ${\bf a}$ for the vacua discussed in Section 2. In $\mathcal{A}_{m}^{(2,1)-}$ the index ${ \bf{a}}$ runs through integer, while in $\mathcal{A}_{m}^{(2,1)+}$ it runs through half-integer values, so it is necessary to consider these subsets separately. It is also important to distinguish whether $m$ is even or odd. 

\begin{itemize}
\item 
For even $m$, expression (\ref{eq:13vevin21}) has a single maximum for both the two subsets of integers and half-integers, lying at ${\bf a}=0$ for the integer and at ${\bf a} = (m-1)/2$ for the half-integer vacua. 
The maximum has equal magnitude in both cases.   Hence,  there is always a single vacuum state in the theory if we approach the horizontal axis from below, independently of whether this is the positive or the negative horizontal axis. 

There is also a single minimum: for the half-integer vacua, this is localized at $\mathbf{a} = m/4$  (if $m = 2 (2 k+1)$), or at $\mathbf{a} = (m-2)/4$ (if $m = 2 (2 k)$); for the integer vacua, it is localized at $\mathbf{a} = (m+4)/4$ (if $m=2 ( 2 k)$) and at $\mathbf{a} = (m+2)/4$ (if $m=2 (2 k+1)$). Hence, there is also a single vacuum state if we approach the negative or the positive horizontal axis from above, although it changes position after crossing the horizontal axis.  Hence, for $m$ even, we refer to Figure \ref{vacuaevolutionn2} for the evolution of the vacuum states. 

\item For odd $m$, the maxima lie at
\begin{eqnarray*}
\mathbf{a}=0\mbox{ and }\mathbf{a}=\frac{m-1}{2} &  & \mbox{for } \mathbf{a}\mbox{ integer}\\
\mathbf{a}=\frac{1}{2}\mbox{ and }\mathbf{a}=\frac{m-2}{2} &  & \mbox{for }\mathbf{a}\mbox{ half-integer}
\end{eqnarray*}
Approaching the horizontal axis from below, 
the theory has two vacua: these are nothing else but the two vacua that were selected out by the perturbation that moves the theory away from negative vertical axis and that change their shape moving toward the horizontal axis. 
 
For the minima, when $m$ is odd, the situation is more articulated, as shown below: in fact, if $\frac{m+1}{2}$ is odd, the minima are at 
\begin{eqnarray*}
\mathbf{a}=\frac{m-1}{4} &  & \mbox{for }\mathbf{a}\mbox{ integer}\\
\mathbf{a}=\frac{m-3}{4} \mbox{ and }\mathbf{a}=\frac{m+1}{4} &  & \mbox{for }\mathbf{a}\mbox{ half-integer}
\end{eqnarray*}
while, if $\frac{m+1}{2}$ is even, the minima are at 
\begin{eqnarray*}
\mathbf{a}=\frac{m-3}{4} \mbox{ and }\mathbf{a}=\frac{m+1}{4} &  & \mbox{for }\mathbf{a}\mbox{ integer}\\
\mathbf{a}=\frac{m-1}{4} &  & \mbox{for }\mathbf{a}\mbox{ half-integer}
\end{eqnarray*}
In the first case ($\frac{m+1}{2}$ odd), moving up from the horizontal axis, there are two vacua in the first quadrant and one vacuum in the third quadrant. In the second case ($\frac{m+1}{2}$ even), the situation is reversed: there is one vacuum in the first quadrant and two vacua in the third quadrant. 
\end{itemize}
To understand the above features, let's examine the adjacency rules in the quantum group picture appropriate to $\Phi_{2,1}$. Now 
\[
j_{\mathrm{max}}=\frac{m-1}{2}
\]
because $q$ is given by (\ref{eq:q21}). The spin of the intertwiner $\varphi_{1,m}$ is exactly $j_{\mathrm{max}}$ and the relevant truncated
fusion rules can be written in the form (\ref{eq:tprules21_13}), which is consistent with the identification of the vacua. Notice that the above positions of the minima can be written as 
\[
\mathbf{a}=\frac{m-1}{4}=\frac{j_{\mathrm{max}}}{2}\quad\mbox{or}\quad \mathbf{a}=\frac{j_{\mathrm{max}}\pm1}{2}
\]
(according to the subset of vacuum labels $\mathbf{a}$ considered) and so depending on the sign of $\Phi_{2,1}$ coupling $\mu$ there are either one or two vacua. It remains to check whether these are consistent with the existence of the topological charge carried by $\varphi_{1,m}$.
Using the truncated quantum group tensor product rules (\ref{eq:rsos_fusion})
\begin{eqnarray*}
j_{\mathrm{max}}\otimes\frac{j_{\mathrm{max}}}{2} & = & \frac{j_{\mathrm{max}}}{2}\\
j_{\mathrm{max}}\otimes\frac{j_{\mathrm{max}}\pm1}{2} & = & \frac{j_{\mathrm{max}}\mp1}{2}
\end{eqnarray*}
so the topological charge of $\varphi_{1,m}$ can be supported by the vacua left after switching on $\lambda>0$ for both even/odd values of $m$.

\begin{figure}[t]
\begin{centering}
\includegraphics[scale=0.4]{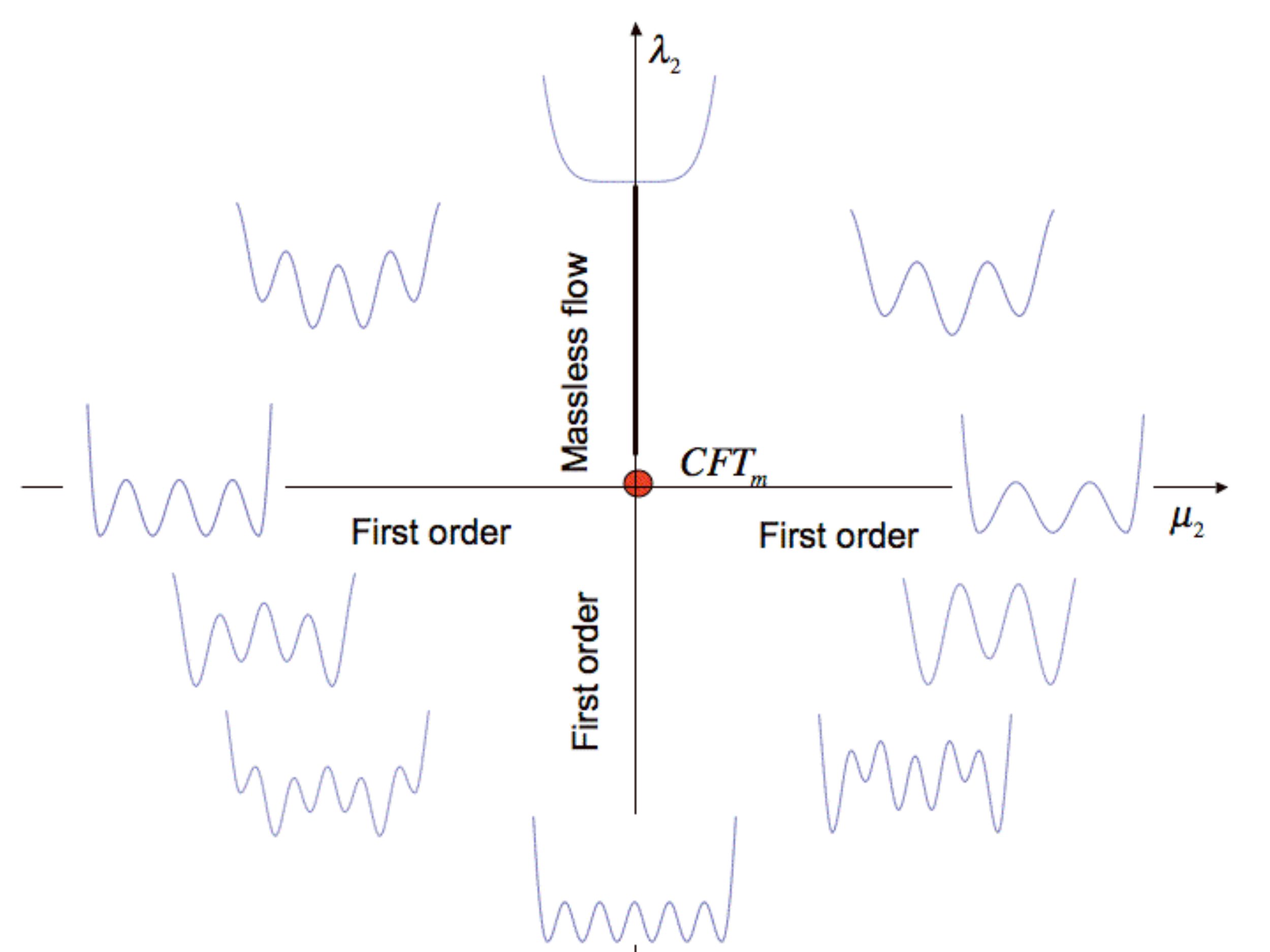}
\par\end{centering}

\caption{\label{vacuaevolution2} Evolution of the vacua structure of the $\mathcal{A}_{m}^{(2)}$ theory in the plane of the couplings when $m$ is odd. This example refers to $m=7$.}
\end{figure}

We can also consider the action of the perturbing operators $\Phi_{1,3}$ and $\Phi_{2,1}$ on the Hilbert space (\ref{eq:topzerospace}). For
an odd value of $m=2k+1$ it turns out that the Hilbert space can be split into two sectors as follows:
\begin{eqnarray*}
\mathcal{H}_{+} & = & \bigoplus_{r=1}^{k}\bigoplus_{s\,\mathrm{odd}}\mathcal{V}_{r,s}\otimes\bar{\mathcal{V}}_{r,s}\\
\mathcal{H}_{-} & = & \bigoplus_{r=1}^{k}\bigoplus_{s\,\mathrm{even}}\mathcal{V}_{r,s}\otimes\bar{\mathcal{V}}_{r,s}
\end{eqnarray*}
while no such splitting can be done for $m$ even, which gives additional support to our conclusions.

\subsection{Perturbing with $\Phi_{1,2}$ and $\Phi_{2,1}$}\label{subsec:1221}

The action is
\begin{equation}
\mathcal{A}_{m}^{(3)}\,=\,
\mathcal{A}_{m}^{(CFT)} - \lambda_3 \int dzd\bar{z}\, \Phi_{1,2}(z,\bar{z}) - \mu_3 \int dzd\bar{z} \,\Phi_{2,1}(z,\bar{z})
\label{eq:action1221}
\end{equation}
Let us consider the model as a perturbation of $\mathcal{A}_{m}^{(1,2)\pm}$ with the operator $\Phi_{2,1}$ (swapping the the roles of the two fields only means interchanging the two integers $m$ and $m+1$ characterizing the minimal model). From eqn. (\ref{exactvev12}) we obtain
\[
\langle {\bf a} | \Phi_{2,1} | {\bf a} \rangle^{(1,2)}\, =\, 2
(-1)^{(2\mathbf{a}+1)} \,\cos\frac{\pi (2\mathbf{a}+1)}{m} \, G_{2,1}^{m}(\lambda_3)
\label{vac12pert21}\]
For $m$ even, all these values are nondegenerate; for $m$ odd they are invariant under the symmetry
\[
\mathbf{a}\,\rightarrow\,j_{\mathrm{max}}-\mathbf{a}\qquad\textrm{where }j_{\mathrm{max}}=\frac{m-2}{2}
\]
but it interchanges integer with half-integer vacua, which cannot be simultaneously present. Therefore the vacuum degeneracy is completely lifted. This is also consistent with the fact that there is no topologically charged operator that can be simultaneously local to both perturbing fields. Furthermore, the fusion rules of the two fields are such that there is no subdivision of the Hilbert space (\ref{eq:topzerospace}) (apart from that corresponding to the conformal spin) which makes the Hamiltonian (\ref{eq:pcftham0}) block-diagonal. Therefore all the kinks are confined and only scalar particles are present in the spectrum.

\section{Application to two-frequency sine-Gordon model}\label{sec:dsg}

An interesting limit of the models considered so far is obtained when $m\rightarrow\infty$. In this limit, the dimension of operator $\Phi_{1,3}$ tends to $1$ and, from the model $\mathcal{A}_{m}^{(1,3)-}$, one obtains the sine-Gordon model\[
\int dzd\bar{z}\,\left(\frac{1}{2}\partial_\mu\phi\partial^\mu\phi+\lambda\cos\beta\phi\right)\]
at the Kosterlitz-Thouless point $\beta=\sqrt{8\pi}$, where it describes an asymptotically free field theory.

Considering now the operators $\Phi_{1,2}$ and $\Phi_{2,1}$, their dimensions tend to $1/4$, and therefore in the limit they can be identified with the operators\[\cos\frac{1}{2}\beta\phi\qquad\mathrm{and}\qquad\sin\frac{1}{2}\beta\phi
\] 
For a more precise description, it is necessary to select the $\mathbb{Z}_2$ symmetry in sine-Gordon theory which has to be identified with the $\mathbb{Z}_2$ transformation of the minimal model. Our convention is to take 
\begin{equation} 
\mathbb{Z}_2:\quad\phi\rightarrow-\phi
\label{choiceZ2} 
\end{equation}
Notice, however, that there are infinitely many other choices related to the above one by the periodicity of the sine-Gordon potential under
\[
\phi\rightarrow\phi+\frac{2\beta}{\pi} \,\,\,, 
\]
each of them corresponding to a convention of how to deal with the zero mode of the bosonic field $\phi$. Within our choice (\ref{choiceZ2}), the even operators ($\Phi_{1,2}$/$\Phi_{2,1}$ for $m$ even/odd, respectively) are identified with $\cos\beta\phi/2$ in the limit, while the odd ones ($\Phi_{2,1}$/$\Phi_{1,2}$ for $m$ even/odd, respectively) are identified with $\sin\beta\phi/2$. Hence, we expect that the limiting models ${\mathcal A}_m^{(1)}$ and ${\mathcal A}_m^{(2)}$ have to do with the non-integrable double sine-Gordon model investigated in \cite{dsg}, whereas the model ${\mathcal A}_m^{(3)}$ with a pure sine-Gordon model but at $\beta^2=2\pi$. Let's investigate in more detail the emergence of these identifications.  

According to our previous results, perturbing $\mathcal{A}_{m}^{(1,3)-}$ with $\Phi_{1,2}$ lifts every second vacuum state. Examining the effective potential of this theory, it is easy to see that the same effect happens in the limit $m\rightarrow\infty$, {\em independently} of whether $m$ is even or odd, i.e. whether the limiting operator is cosine or sine. This also means that the same must be true for $\Phi_{2,1}$. But there is an apparent obstacle to this conclusion, which comes from the behavior 
at finite $m$ of the models ${\mathcal A}^{(2)}_m$,  where there is the simultaneous presence of the fields $\Phi_{1,3}$ and $\Phi_{2,1}$: these models, in fact, have only one or two vacua left (depending on the parity of $m$). How to handle this paradox? The solution is in the limiting values of the vacuum expectation values: note that the vacuum shifts given in eqn. (\ref{vac13pert21}) satisfy
\[
\delta\mathcal{E}_\mathbf{a}\propto (-1)^{2\mathbf{a}+1}\cos\frac{\pi(2\mathbf{a}+1)}{m}\ \mathop{\longrightarrow}_{m\rightarrow\infty}\ (-1)^{2\mathbf{a}+1}
\]
and so they become equal in magnitude while alternating in sign, consistent with the limiting double sine-Gordon picture. A further check that this is indeed the solution of the paradox is to consider the picture in which the perturbing operator is $\Phi_{1,3}$ added to the integrable theory defined by $\Phi_{2,1}$: in this case, from the limiting sine-Gordon theory, we expect that none of the vacua permitted by $\Phi_{2,1}$ will be lifted in the limit. According to eqn. (\ref{eq:13vevin21}) the vacuum shifts are
\[
\delta\mathcal{E}_\mathbf{a}\propto \frac{\sin\left(\frac{3\pi(2\mathbf{a}+1)}{m+1}\right)}{\sin\frac{\pi(2\mathbf{a}+1)}{m+1}}
\]
and the right hand side expression tends to the constant value $3$ in the limit, i.e. to a uniform shift of {\em all} vacua that correctly does not lift any of them!  

For the model $\mathcal{A}_{m}^{(3)}$ the situation is even more interesting. The sine-Gordon potential in this case is expected to be 
\[
\lambda_1\cos\frac{1}{2}\beta\phi+\lambda_2\sin\frac{1}{2}\beta\phi =\lambda\cos\frac{1}{2}\beta(\phi-\phi_0) 
\]
with 
\[
\lambda=\sqrt{\lambda_1^2+\lambda_2^2}\qquad,\qquad\tan\frac{1}{2}\beta\phi_0=\frac{\lambda_2}{\lambda_1}
\]
\begin{figure}[t]
\begin{centering}
\includegraphics[scale=0.35]{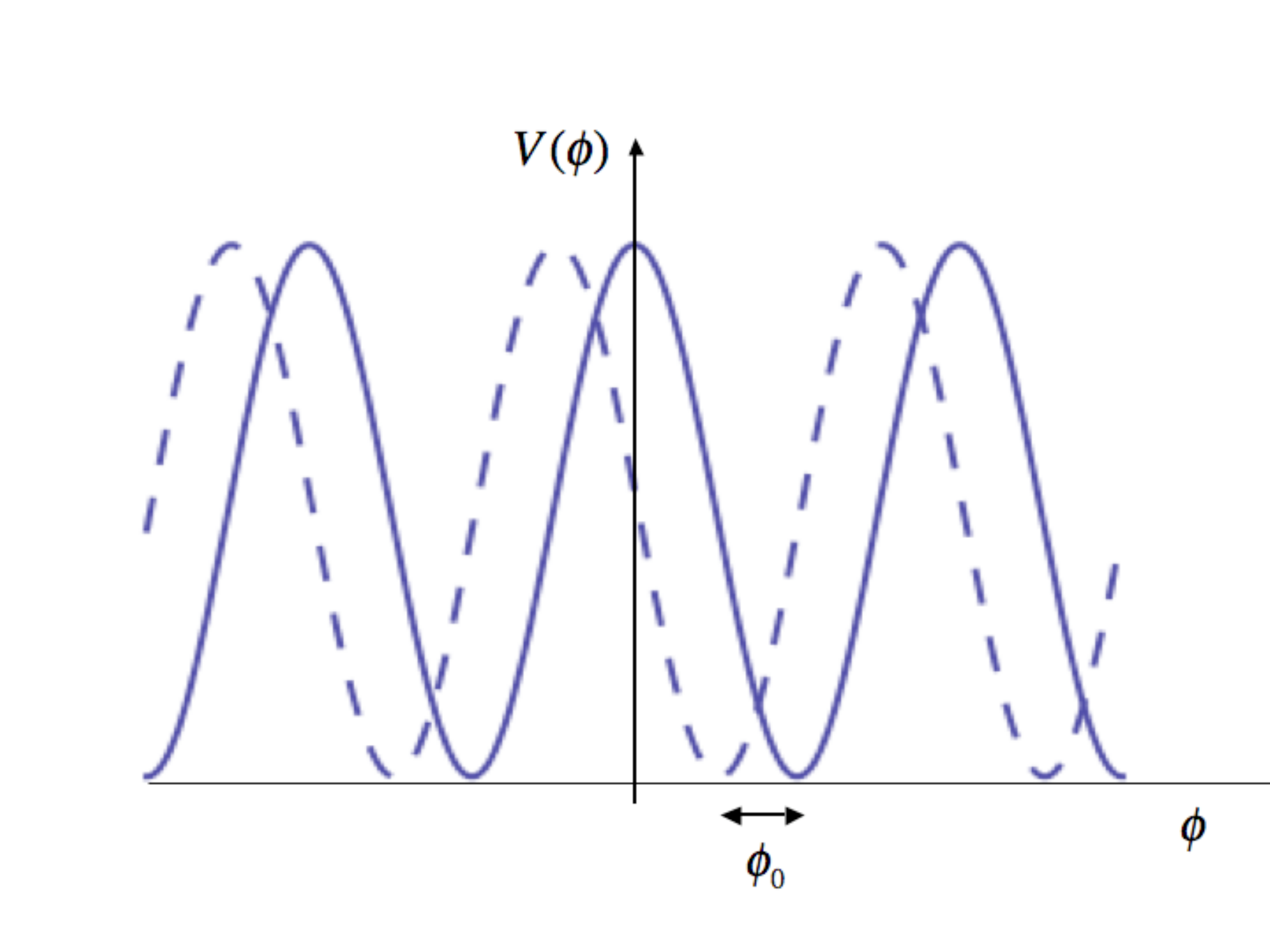}
\par\end{centering}

\caption{\label{shiftpot} Original sine-Gordon potential (continuous line) and shifted potential (dashed line) under deformation.}
\end{figure}
If this conclusion is correct, it means that the degeneracy of the original vacua should not be lifted by adding the other perturbation: the only effect should simply be a translation of the original effective potential by the amount $\phi_0$ (see Figure \ref{shiftpot}).  This picture seems to be in contrast with the one coming from the perturbed minimal models, where we saw that the double perturbation selected instead a unique ground state. The solution of this paradox is once again in the limiting values of the vacuum expectation values: eqn. (\ref{vac12pert21}) gives 
\[
\delta\mathcal{E}_\mathbf{a}\propto (-1)^{2\mathbf{a}+1}\cos\frac{\pi(2\mathbf{a}+1)}{m}\ \mathop{\longrightarrow}_{m\rightarrow\infty}\ (-1)^{2\mathbf{a}+1}
\]
which indicates that in the limit $m \rightarrow \infty$, the vacuum energy shifts become equal in magnitude, but alternating in sign. But we must remember that in the model $\mathcal{A}_{m}^{(1,2)\pm}$ only the integer/half-integer vacua are present (depending on the sign of the coupling), and so we obtain consistency with the sine-Gordon picture as before. Furthermore, since in the limiting theory the effect of the perturbation (apart from redefining the coupling and thus also the mass scale) is simply a uniform displacement of the vacua by $\phi_0$, the original kinks must be present also in the perturbed theory, without feeling any confinement. This can be easily proved. From the point of view of the minimal model, note that the chiral operators $\varphi_{2k+1,1}$ creating the topologically charged sectors of the theory $\mathcal{A}_{m}^{(1,2)\pm}$ become local with respect to the perturbation $\Phi_{2,1}$ in the limit $m\rightarrow\infty$, and this agrees with the fact that all the solitonic excitations survive.

Finally we remark that interchanging the roles of the two fields $\Phi_{1,2}$ and $\Phi_{2,1}$, the above considerations can be applied again simply by swapping the roles of the minimal model indices $m$ and $m+1$.

\section{Conclusions}\label{sec:conclusions}
In this paper, using simple arguments based on form factor perturbation theory, exact expressions of vacuum expectation values and truncated conformal space approach, we establish the evolution of the effective potential in certain non-integrable deformations of minimal models. For small values of the coupling involved, the analysis is both qualitative and quantitative, while for intermediate values of the couplings, it only provides a qualitative scenario of the corresponding field theory. Nevertheless this  information is useful because it correctly identifies the set of stable and unstable vacua, together with the nature of the massive excitations above them:  these data are important as the starting point of a more refined analysis of these theories. We have also investigated the interesting scenario that emerges in the limit $m\rightarrow \infty$, with a relevant role played by the Sine-Gordon model and deformations thereof. 

\subsection*{Acknowledgments}
We would like to thank the Asian Pacific Center of Theoretical Physics
in Pohang and the Galileo Galilei Institute in Florence where this
work was started and completed. G.T. also thanks SISSA and G. Mussardo
for their warm hospitality. This work was partially supported by the
ESF grant INSTANS, the MIUR grant {\em Fisica Statistica dei Sistemi
  Fortemente Correlati all'Equilibrio e Fuori Equilibrio: Risultati
  Esatti e Metodi di Teoria dei Campi} and by the Hungarian OTKA
grants K60040 and K75172.

%\newpage
\vspace{3mm}
\appendix
\makeatletter 
\renewcommand{\theequation}{\hbox{\normalsize\Alph{section}.\arabic{equation}}} 
\@addtoreset{equation}{section} 
\renewcommand{\thefigure}{\hbox{\normalsize\Alph{section}.\arabic{figure}}} 
\@addtoreset{figure}{section} 
\renewcommand{\thetable}{\hbox{\normalsize\Alph{section}.\arabic{table}}} 
\@addtoreset{table}{section}
\makeatother

\section{The models $\mathcal{A}_{6}^{(1,2)\pm}$}

\begin{figure}
\begin{centering}
\psfrag{mb1}{$\scriptstyle m_{1}$}
\psfrag{mb2}{$\scriptstyle m_{2}$}
\psfrag{mk1}{$\scriptstyle M_{1}$}
\psfrag{mk2}{$\scriptstyle M_{2}$}
\psfrag{l}{$l$}
\psfrag{e(l)}{$e(l)$}\subfigure[Sector $\mathcal{H}_{1}$ of $\mathcal{A}_6^{ (1,2)\pm }$]{\includegraphics[scale=0.6]{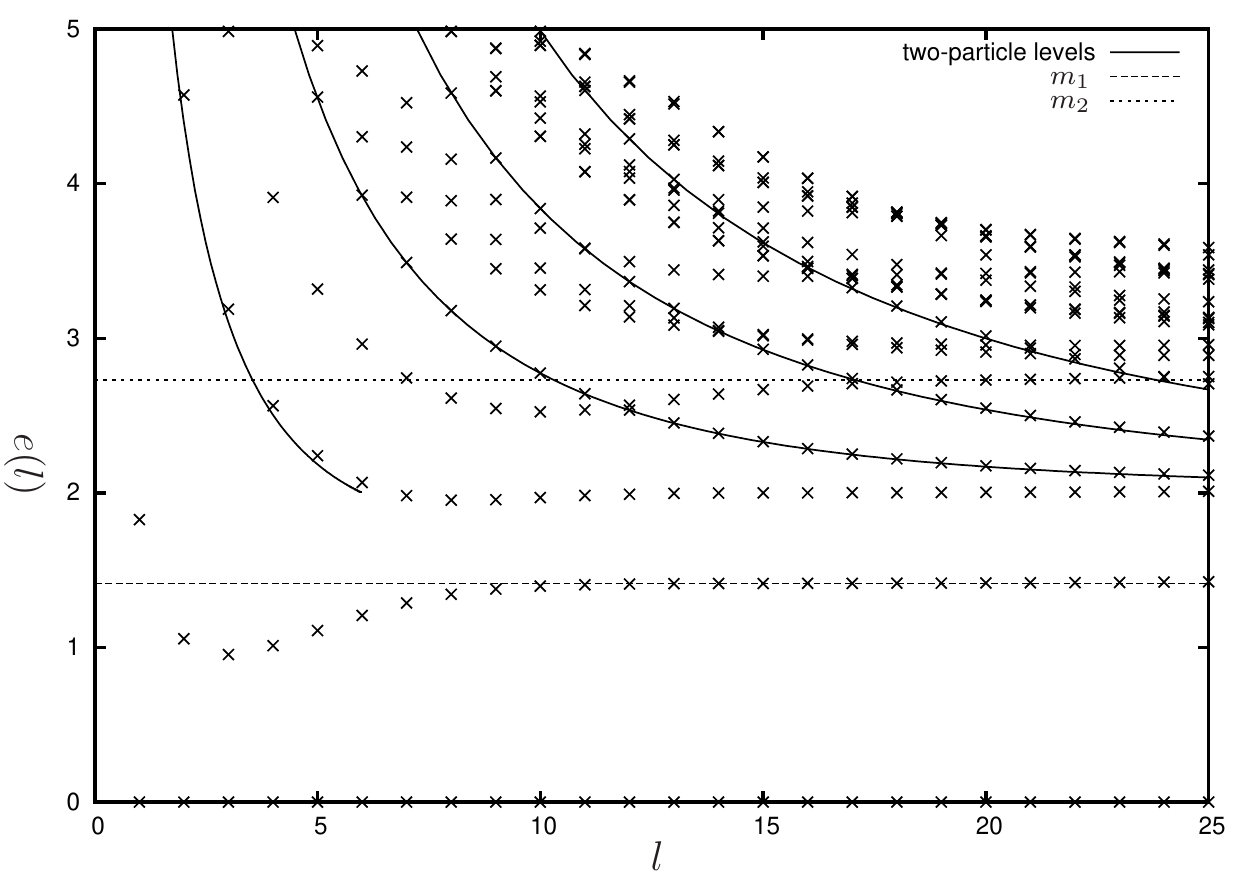}}
\psfrag{mb1}{$\scriptstyle m_{1}$}
\psfrag{mb2}{$\scriptstyle m_{2}$}
\psfrag{mk1}{$\scriptstyle M_{1}$}
\psfrag{mk2}{$\scriptstyle M_{2}$}
\psfrag{l}{$l$}
\psfrag{e(l)}{$e(l)$}\subfigure[Sector $\mathcal{H}_{2}$ of $\mathcal{A}_6^{ (1,2)\pm }$]{\includegraphics[scale=0.6]{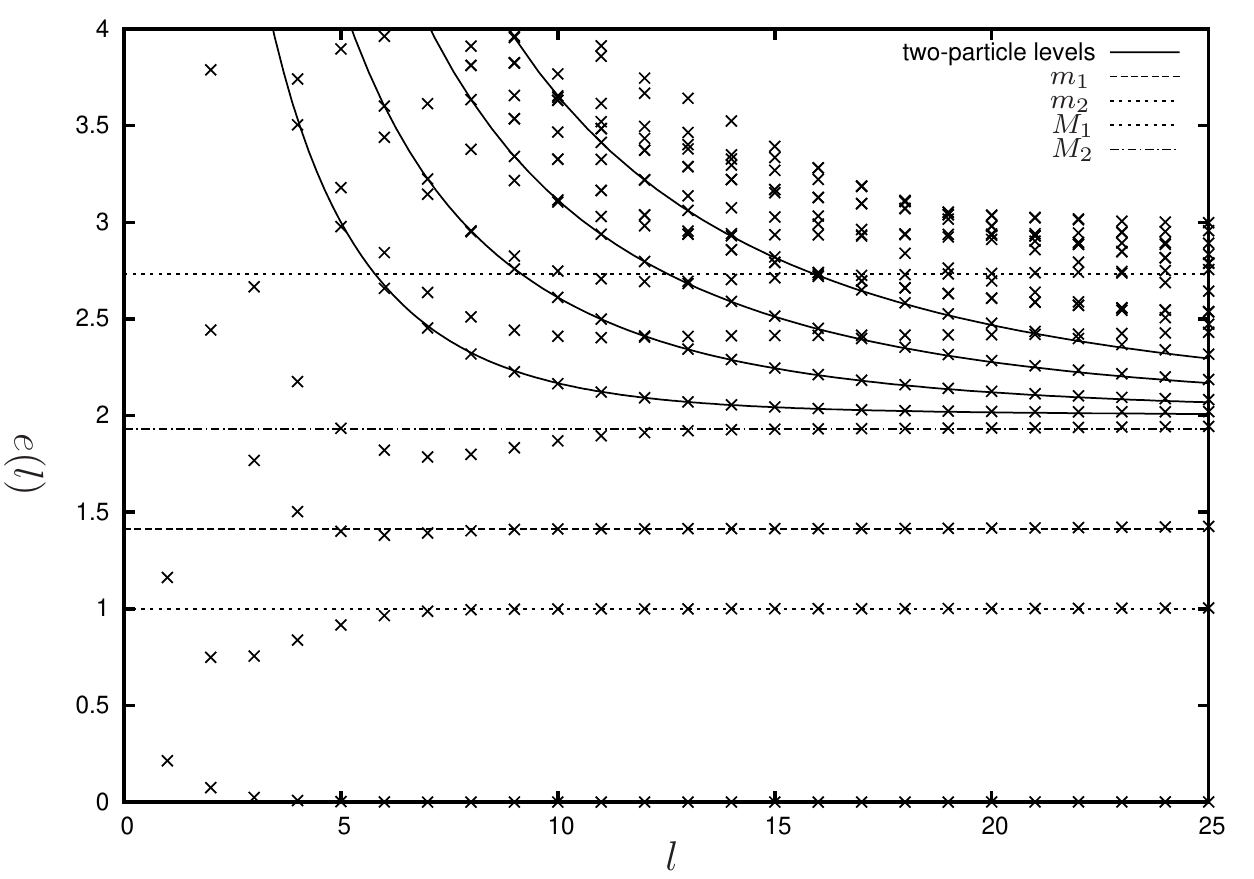}}
\par\end{centering}

\caption{\label{fig:tcsaeven} The spectrum of spin-zero states in sectors
$\mathcal{H}_{1}$ and $\mathcal{H}_{2}$ for the models $\mathcal{A}_{6}^{(1,2)\pm}$.
The number of states kept are $2412$ and $5987$, corresponding to
a cutoff at level $20$ for both sectors.}

\end{figure}

\begin{figure}
\begin{centering}
\psfrag{mb1}{$\scriptstyle m_{1}$}
\psfrag{mb2}{$\scriptstyle m_{2}$}
\psfrag{mk1}{$\scriptstyle M_{1}$}
\psfrag{mk2}{$\scriptstyle M_{2}$}
\psfrag{l}{$l$}
\psfrag{e(l)}{$e(l)$}\subfigure[Sector $\mathcal{H}_{3}$ of $\mathcal{A}_6^{ (1,2) + }$]{\includegraphics[scale=0.6]{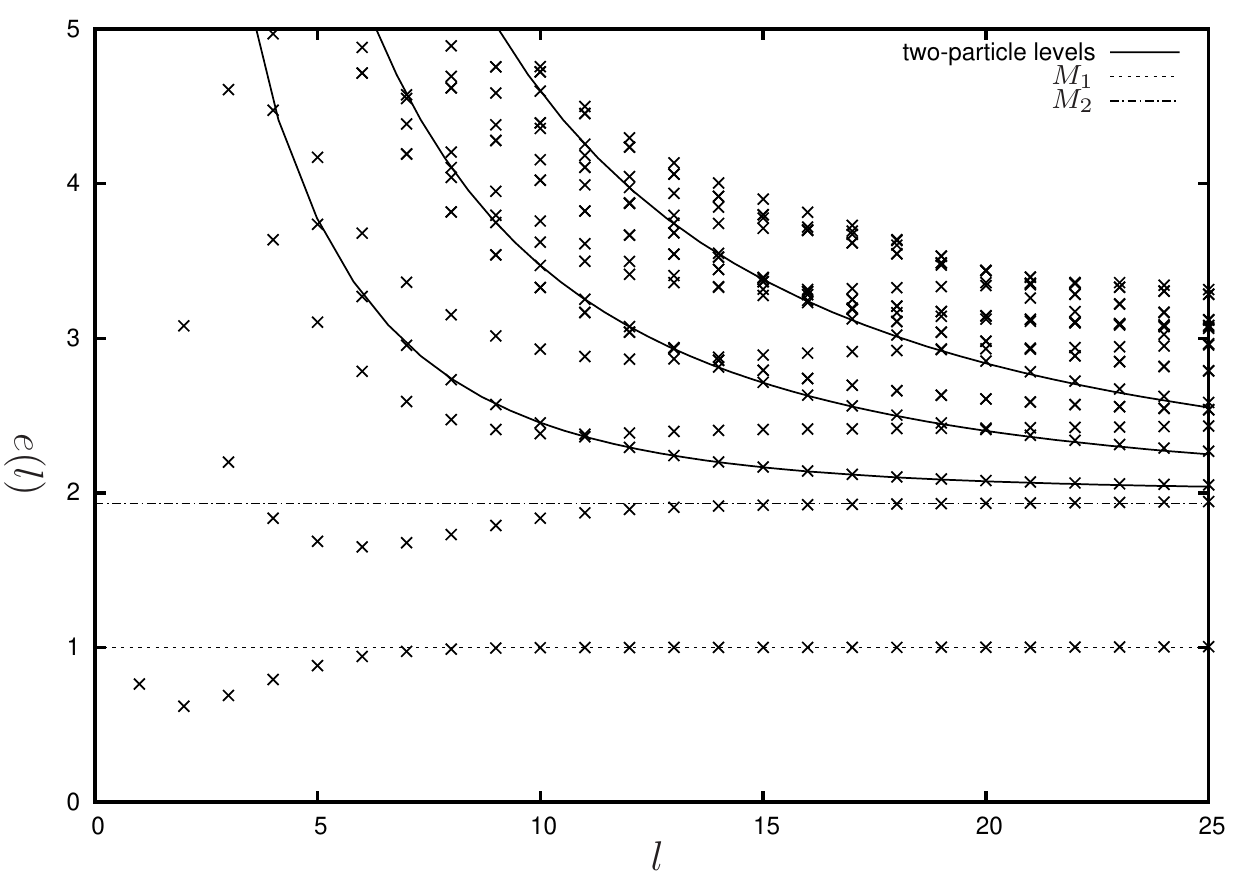}}
\psfrag{mb1}{$\scriptstyle m_{1}$}
\psfrag{mb2}{$\scriptstyle m_{2}$}
\psfrag{mk1}{$\scriptstyle M_{1}$}
\psfrag{mk2}{$\scriptstyle M_{2}$}
\psfrag{l}{$l$}
\psfrag{e(l)}{$e(l)$}\subfigure[Sector $\mathcal{H}_3$ of $\mathcal{A}_6^{ (1,2) - }$]{\includegraphics[scale=0.6]{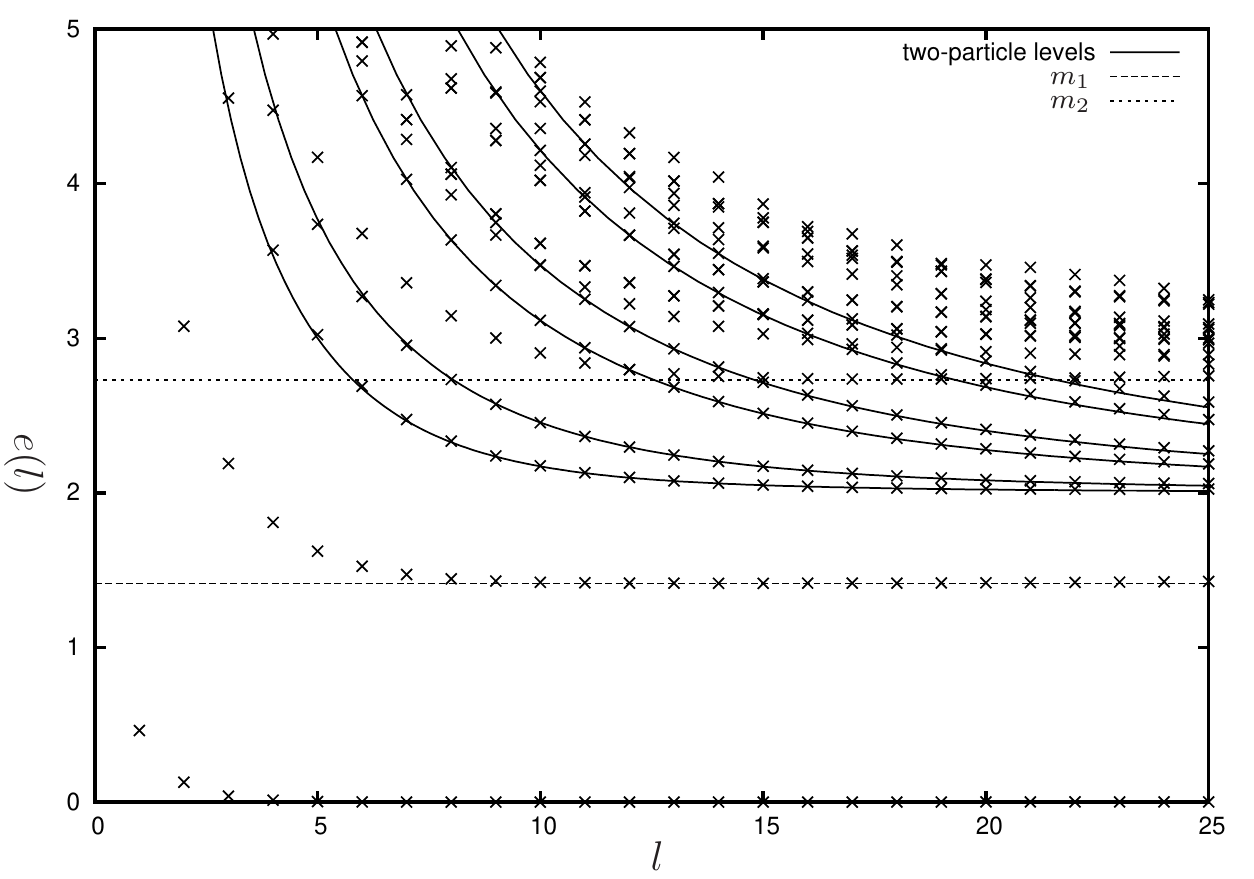}}
\par\end{centering}

\caption{\label{fig:tcsaodd} The spectrum of spin-zero states in sector $\mathcal{H}_{3}$
for the models $\mathcal{A}_{6}^{(1,2)\pm}$. The number of states
kept is $3714$, corresponding to a cutoff at level $20$.}

\end{figure}

The mass $M_{1}$ of fundamental kink   can be expressed as a function of the coupling \cite{massgaps}
\[
\lambda=0.10390339\dots\times M_{1}^{12/7}
\]
This relation can be used to express the Hamiltonian (\ref{eq:pcftham})
in terms of dimensionless energy and volume variables\[
e=E/M_{1}\quad,\quad l=M_{1}L\]
There is also an excited kink with mass \[
M_{2}=2M_{1}\sin\left(\frac{\pi}{6}+\frac{\xi}{2}\right)=1.93185\dots\times M_{1}\quad,\quad\xi=\frac{\pi}{2}\]
and two breathers with masses
\[
m_{1}=2M_{1}\sin\frac{\xi}{2}=1.4141\dots\times M_{1}\quad,\quad m_{2}=4M_{1}\sin\left(\frac{\pi}{6}+\frac{\xi}{2}\right)\sin\frac{\xi}{2}=2.73205\dots\times M_{1}
\]
The spectrum can be evaluated using the truncated conformal approach
\cite{tcsa} and the results are presented in the figures \ref{fig:tcsaeven}
and \ref{fig:tcsaodd}. We indicated the predicted masses in the plots;
they are reproduced up to deviations of order $10^{-3}$.

To distinguish between the two scattering theories, one based on the
integer and the other on the half-integer vacua, it is necessary to
examine the two-particle levels. They can be predicted from the exact
$S$ matrix using the transfer matrix formalism developed in \cite{finiteVkinks}
(see also \cite{rsos} for more details). The two-kink states (of zero total momentum) take the form\[
|K_{ab}(\theta)K_{bc}(-\theta)\rangle\]
with periodic boundary conditions demanding that $a=c$. For integer
vacua there are $5$ such states allowed by the adjacency conditions:\[
|K_{01}(\theta)K_{10}(-\theta)\rangle,\;|K_{10}(\theta)K_{01}(-\theta)\rangle,\;|K_{11}(\theta)K_{11}(-\theta)\rangle,\;|K_{12}(\theta)K_{21}(-\theta)\rangle,\;|K_{21}(\theta)K_{12}(-\theta)\rangle\]
while for half-integer vacua we find $4$ states \[
|K_{\frac{1}{2}\frac{1}{2}}(\theta)K_{\frac{1}{2}\frac{1}{2}}(-\theta)\rangle,\;|K_{\frac{1}{2}\frac{3}{2}}(\theta)K_{\frac{3}{2}\frac{1}{2}}(-\theta)\rangle,\;|K_{\frac{3}{2}\frac{1}{2}}(\theta)K_{\frac{1}{2}\frac{3}{2}}(-\theta)\rangle,\;|K_{\frac{3}{2}\frac{3}{2}}(\theta)K_{\frac{3}{2}\frac{3}{2}}(-\theta)\rangle\]
The corresponding transfer matrices have the following eigenvalues:
\begin{eqnarray}
\Lambda_{k}^{(-)}(\theta) & = & \mathrm{e}^{i(2k+1)\frac{\pi}{3}}\sinh\left(2\theta\right)\sinh\left(2\theta-i\frac{\pi}{3}\right)\sigma(\theta)\quad,k=1,2,3 \nonumber\\
\Lambda_{4}^{(-)}(\theta) & = & -\Lambda_{5}^{(-)}(\theta)=\frac{1}{2}\sinh\left(2\theta\right)\sqrt{1+2\cosh4\theta}\sigma(\theta)\label{minusevals}
\end{eqnarray}
for the integer case, and
\begin{eqnarray}
\Lambda_{1}^{(+)}(\theta) & = & \Lambda_{1}^{(-)}(\theta)\quad,\quad\Lambda_{2}^{(+)}(\theta)=\Lambda_{4}^{(-)}(\theta)\quad,\quad\Lambda_{3}^{(+)}(\theta)=\Lambda_{5}^{(-)}(\theta)=-\Lambda_{4}^{(-)}(\theta) \nonumber\\
\Lambda_{4}^{(+)}(\theta) & = & -\Lambda_{1}^{(-)}(\theta)\label{plusevals}
\end{eqnarray}
for the half-integer case, where
\[
\sigma(\theta)=\frac{\exp\left(-2i\int_{0}^{\infty}\frac{dk}{k}\sin k\theta\frac{\sinh\frac{\pi k}{3}\cosh\left(\frac{\pi}{6}-\frac{\xi}{2}\right)k}{\cosh\frac{\pi k}{2}\sinh\frac{\xi k}{2}}\right)}{\sinh\frac{\pi}{\xi}(\theta-i\pi)\sinh\frac{\pi}{\xi}\left(\theta-\frac{2i\pi}{3}\right)}
\]
Two particle levels can be obtained by solving the Bethe-Yang equation\[
l\sinh\theta+\Lambda(2\theta)=2n\pi\]
where $\Lambda$ is an eigenvalue of the transfer matrix.%

Let's first analyze the data in figure \ref{fig:tcsaeven}. The sectors $\mathcal{H}_{1}$ and $\mathcal{H}_{2}$ are independent of the sign
of coupling. The energy levels were normalized by subtracting the
ground state of $\mathcal{H}_{1}$, and it is apparent that both sectors
contain a vacuum state. In addition, there is a one-kink state in
$\mathcal{H}_{2}$, corresponding to a link in the adjacency diagram
that connects a vacuum with itself. The two-particle states of these
sectors can be explained by both the integer and the half-integer
transfer matrix: the ones in $\mathcal{H}_{1}$ are described by the
eigenvalue $\Lambda_{1}^{(-)}=\Lambda_{1}^{(+)}$, while those in
$\mathcal{H}_{2}$ correspond to $\Lambda_{2}^{(+)}=\Lambda_{4}^{(-)}$
and $\Lambda_{3}^{(+)}=\Lambda_{5}^{(-)}$. We remark that the apparently
incomplete two-particle level in sector $\mathcal{H}_{1}$ corresponds
to a would-be second breather bound state of fundamental kinks which
is exactly on threshold in infinite volume, but becomes bound by finite
size effects exponentially decaying in the volume. At a certain critical
volume its energy crosses the threshold $2M_{1}$ again, and for smaller
values of $l$ it can be described as a two-kink state which is shown
as the incomplete continuous line.

From figure \ref{fig:tcsaodd} it is obvious that the spectrum of
sector $\mathcal{H}_{3}$ does depend on the sign of the coupling.
For $\lambda>0$ (model $\mathcal{A}_{6}^{(1,2)+}$) there is no vacuum
state in this sector, but we find an additional neutral kink state.
This is in accordance with the adjacency diagram for half-integer
vacua: we expect only two vacua $1/2$ and $3/2$, and both of them
should have neutral kink excitations over them. In accordance with
this, that the two-particle levels in this sector turn out to correspond
to the eigenvalue $\Lambda_{4}^{(+)}$ which is only present for the
half-integer transfer matrix. On the other hand, for $\lambda<0$
(model $\mathcal{A}_{6}^{(1,2)-}$) there is an additional vacuum,
but not additional neutral kink state, which fits well with the expectation
that there must be three vacua $0$, $1$ and $2$, but only $1$
can support neutral kink excitations. In addition, the two-particle
levels are now described by the eigenvalues $\Lambda_{2}^{(-)}$ and
$\Lambda_{3}^{(-)}$.

One can calculate the lowest energy level in $\mathcal{H}_{3}$ using
perturbation theory. To first order, the result is\[
E_{\pm}(L)=\frac{2\pi}{L}\left(2h_{33}-\frac{c}{12}\pm\frac{\lambda L^{2-2h_{1,2}}}{(2\pi)^{1-2h_{1,2}}}C_{(1,2)(3,3)}^{(3,3)}+O(\lambda^{2})\right)\]
Using the normalization condition (\ref{eq:sign_convention}) we see
that it receives a large (perturbative) correction which is positive
for $\mathcal{A}_{6}^{(1,2)+}$ and negative for $\mathcal{A}_{6}^{(1,2)-}$
. According to the arguments presented in section 2.2 
(in the paragraph preceding eqn. (\ref{eq:12sectorseven})), 
this indicates that the difference $E_{+}(L)-E_{-}(L)$
has a finite positive limit when $L\rightarrow\infty$. Since it is
expected that exactly one of these states is a vacuum state, while
the other one has a finite gap over the vacuum, it follows that $\mathcal{A}_{6}^{(1,2)-}$
has three ground states while $\mathcal{A}_{6}^{(1,2)+}$only has
two, which is consistent with the proposition made at the end of subsection
3.1. Indeed, the above argument can be extended to the general case
$\mathcal{A}_{2k}^{(1,2)\pm}$, which justifies the proposition at
the end of subsection 3.1.

\section{Exact vacuum expectation values}

In this appendix we collect for convenience the formulae for the exact vacuum expectation
values of primary fields in perturbed conformal field theories, derived in \cite{exactvev}. Hereafter \[
x\equiv (m+1) k -m l\]

\vspace{3mm}
\noindent
{\bf Integrable theory $\mathcal{A}_{m}^{(1,3)-}$}

\noindent
In the case of the model $\mathcal{A}_{m}^{(1,3)-}$ for the vacuum expectation values of the primary fields on the various vacua we have 
\begin{equation}
\langle {\bf a} |\Phi_{k,l}|{\bf a}\rangle^{(1,3)-}\, =\, 
\frac{\sin\left(\frac{\pi(2a+1)}{m}((m+1)k-ml)\right)}{\sin\frac{\pi(2a+1)}{m}}\, F_{k,l}^{m}(\lambda)\label{app:exactvev13}
\end{equation}
where
\begin{equation}
 F_{k,l}^{m}(\lambda)\,=\,\left( M\frac{\sqrt{\pi}\Gamma\left(\frac{m+3}{2}\right)}{2\Gamma\left(\frac{m}{2}\right)}\right)^{2\Delta_{k,l}}\mathcal{Q}_{1,3}\left(x\right)
\end{equation}
with 
\begin{equation}
\mathcal{Q}_{1,3}(\eta)=\exp\left\{
\int_0^\infty\frac{dt}{t}\left[
\frac{\cosh(2t)\sinh((\eta-1)t)\sinh((\eta+1)t)}{2\cosh(t)\sinh(mt)\sinh((1+m)t)}
-\frac{\eta^2-1}{2m(m+1)}\mathrm{e}^{-4t}\right] 
\right\}
\end{equation}
In the formula above 
\begin{equation}
M\,=\,\frac{2\Gamma\left(\frac{m}{2}\right)}{\sqrt{\pi}\Gamma\left(\frac{m+1}{2}\right)}
\left[\frac{\pi\lambda(1-m)(2m-1)}{(1+m)^2}
\sqrt{\frac{\Gamma\left(\frac{1}{m+1}\right)\Gamma\left(\frac{1-2m}{m+1}\right)}
{\Gamma\left(\frac{m}{m+1}\right)\Gamma\left(\frac{3m}{m+1}\right)}}
\right]^{\frac{1+m}{4}}
\end{equation}
is the mass of the kinks expressed in term of the coupling constant $\lambda$.

\vspace{3mm}
\noindent
{\bf Integrable theory $\mathcal{A}_{m}^{(1,2)}$}

\noindent
For the integrable model $\mathcal{A}_{m}^{(1,2)}$ the vacuum expectation values of the primary fields on the various vacua are: 
\begin{equation}
\langle {\bf a} | \Phi_{k,l} | {\bf a} \rangle^{(1,2)}\,=\,
\frac{\sin\left(\frac{\pi(2a+1)}{m}((m+1)k-ml)\right)}{\sin\frac{\pi(2a+1)}{m}} \, 
G_{k,l}^{m}(\lambda)
\label{app:exactvev12}
\end{equation}
where
\begin{equation}
 G_{k,l}^{m}(\lambda)\,=\,\left( M\frac{\pi (m+1)\Gamma\left(\frac{2m+2}{3m+6}\right)}{2^{\frac{2}{3}}\sqrt{3}\Gamma\left(\frac{1}{3}\right)\Gamma\left(\frac{m}{3m+6}\right)}\right)^{2\Delta_{k,l}}\mathcal{Q}_{1,2}\left(x\right)
\end{equation}
with 
\begin{eqnarray}
\mathcal{Q}_{1,2}(\eta)=&&\exp\Bigg\{
\int_0^\infty\frac{dt}{t}\Big[
\frac{\sinh((m+2)t)\sinh((\eta-1)t)\sinh((\eta+1)t)}{\sinh(3(m+2)t)\sinh(2(m+1)t)\sinh(mt)}
\times\nonumber\\
&&\left(
\cosh(3(m+2)t)+\cosh((m+4)t)-\cosh((3m+4)t)+\cosh(mt)+1
\right)
\nonumber\\
&&-\frac{\eta^2-1}{2m(m+1)}\mathrm{e}^{-4t}\Big] 
\Bigg\}
\end{eqnarray}
In the formula above 
\begin{equation}
M\,=\,\frac{2^{\frac{m+5}{3m+6}}\sqrt{3}\Gamma\left(\frac{1}{3}\right)
\Gamma\left(\frac{m}{3m+6}\right)}
{\pi\Gamma\left(\frac{2m+2}{3m+6}\right)}
\left[
\frac{\pi^2\lambda^2\Gamma\left(\frac{3m+4}{4m+4}\right)
\Gamma\left(\frac{1}{2}+\frac{1}{m+1}\right)}
{\Gamma\left(\frac{m}{4m+4}\right)\Gamma\left(\frac{1}{2}-\frac{1}{m+1}\right)}
\right]^{\frac{m+1}{3m+6}}
\end{equation}
is the mass of the kinks expressed in terms of the coupling constant $\lambda$.

\vspace{3mm}

\noindent{\bf Integrable theory $\mathcal{A}_{m}^{(2,1)}$}

\noindent
In this integrable model the vacuum expectation values of the primary fields on the various vacua are  
\begin{equation}
\langle {\bf a} | \Phi_{k,l} | {\bf a} \rangle^{(2,1)}\,=\,
\frac{\sin\left(\frac{\pi(2a+1)}{m+1}((m+1)k-ml)\right)}{\sin\frac{\pi(2a+1)}{m+1}} \, 
H_{k,l}^{m}(\lambda)
\label{app:exactvev21}
\end{equation}
where
\begin{equation}
 H_{k,l}^{m}(\lambda)\,=\,\left( M\frac{\pi m\Gamma\left(\frac{2m}{3m-3}\right)}{2^{\frac{2}{3}}\sqrt{3}\Gamma\left(\frac{1}{3}\right)\Gamma\left(\frac{m+1}{3m-3}\right)}\right)^{2\Delta_{k,l}}\mathcal{Q}_{2,1}\left(x\right)
\end{equation}
with 
\begin{eqnarray}
\mathcal{Q}_{2,1}(\eta)=&&\exp\Bigg\{
\int_0^\infty\frac{dt}{t}\Big[
\frac{\sinh((m-1)t)\sinh((\eta-1)t)\sinh((\eta+1)t)}{\sinh(3(m-1)t)\sinh(2mt)\sinh((m+1)t)}
\times\nonumber\\
&&\left(
\cosh(3(m-1)t)+\cosh((m-3)t)-\cosh((3m-1)t)+\cosh((m+1)t)+1
\right)
\nonumber\\
&&-\frac{\eta^2-1}{2m(m+1)}\mathrm{e}^{-4t}\Big] 
\Bigg\}
\end{eqnarray}
The mass of the kink, as a function of the coupling constant $\lambda$, is expressed by 
\begin{equation}
M\,=\,\frac{2^{\frac{m-4}{3m-3}}\sqrt{3}\Gamma\left(\frac{1}{3}\right)
\Gamma\left(\frac{m+1}{3m-3}\right)}
{\pi\Gamma\left(\frac{2m}{3m-3}\right)}
\left[
\frac{\pi^2\lambda^2\Gamma\left(\frac{3m-1}{4m}\right)
\Gamma\left(\frac{1}{2}-\frac{1}{m}\right)}
{\Gamma\left(\frac{m+1}{4m}\right)\Gamma\left(\frac{1}{2}+\frac{1}{m}\right)}
\right]^{\frac{m}{3m-3}}
\end{equation}

%\newpage 

\end{document}